\PassOptionsToPackage{hyphens}{url}

\documentclass[aps,pra,twocolumn,superscriptaddress,showkeys,floatfix]{revtex4-2}

\usepackage{graphicx}
    \graphicspath{{manuscript_figs/}} 
\usepackage{dcolumn}
\usepackage{bm}
\usepackage{xcolor}
\usepackage{amsfonts}
\usepackage{amssymb}
\usepackage{amsmath}
\usepackage[caption=false]{subfig} 
\usepackage{siunitx} 
\usepackage{float} 
\usepackage{adjustbox}
\usepackage{xfrac} 
\usepackage[english]{babel}
\usepackage{hyperref}
\usepackage{breakurl}

\begin{document}


\title{A Machine-Learning-Accelerated Quantum Transport Study on the Effects of Superlattice Disorder and Strain in a Mid-wave Infrared Curved Sensor}

\author{John Glennon}
 \email{jglennon@bu.edu}
\affiliation{Division of Material Science and Engineering, Boston University, Boston,
MA 02215, USA}
\author{Alexandros Kyrtsos}
\affiliation{Silvaco, Inc., 4701 Patrick Henry Drive, Santa Clara, CA 95054, USA}
\affiliation{Department of Electrical and Computer Engineering, Boston University, Boston, MA 02215, USA}
\author{Mark R. O'Masta}
\affiliation{HRL Laboratories, LLC, 3011 Malibu Canyon Road, Malibu, CA 90265, USA}
\author{Binh-Minh Nguyen}
\affiliation{HRL Laboratories, LLC, 3011 Malibu Canyon Road, Malibu, CA 90265, USA}
\author{\\Enrico Bellotti}
\affiliation{Division of Material Science and Engineering, Boston University, Boston,
MA 02215, USA}
\affiliation{Department of Electrical and Computer Engineering, Boston University, Boston,
MA 02215, USA}

\date{\today}

\begin{abstract}
An emerging device architecture for infrared imaging is the curved focal-plane array which benefits from several optical advantages over the traditional flat design. However, the curving process introduces additional strain in the active region which must be taken into account. Type-II superlattices, a promising alternative to traditional bulk materials for use in infrared photodetectors, is a candidate material for use in these devices, but the transport properties of these highly heterogeneous materials are not straightforward and can be affected by different material conditions, such as superlattice disorder and external strain. We present a comprehensive study of the internal QE calculated for a curved device that incorporates finite element analysis (FEA) modeling, nonequilibirium Green's functions (NEGF) calculations, and Gaussian Process (GP) regression. FEA is used for predicting the strain configuration throughout the active region induced by the curving procedure of the device. NEGF is used to calculate the vertical hole mobility for a select set of strain configurations, from which the internal quantum efficiency of the device is approximated to predict performance under strained conditions. Then this data set is used to train a GP model that maps the quantum efficiency QE predictions onto the spatial coordinates of the curved device, based on the strain configuration predicted using FEA. This analysis is performed for ideal and disordered SLs to understand both the fundamental and practical limitations of the performance of these materials in curved devices.
\end{abstract}

\maketitle

\section{\label{sec:I}Introduction\protect}
An emerging device architecture for the improvement of imaging sensors is the curved focal-plane array (FPA)\,\cite{2008RimOPG}. The motivation for this design comes from the desire to reduce the complexity of the optical system. Traditional flat FPA architectures require large optical systems to achieve field flattening while minimizing aberrations that effect image quality\,\cite{2016RogalskiIOP}. However, utilizing curved FPA architectures can remove the need for field flattening leading to simpler optical designs. Recently, the development of a method for curving photographic sensors in visible-light cameras has resulted in a curved-FPA with improved resolution and nearly uniform illumination compared with professional camera systems with similar lenses, while reducing the size, weight, and cost of the optical system\,\cite{2017GuenterOPG}. This method is applicable to IR FPAs as well and can be expected to similarly improve device performance\,\cite{2022OMastaITA}. However, it is important to understand the effect that curving has on the underlying material properties. Curving the FPA will introduce additional strain throughout the material structure which can affect carrier transport. A recent study has investigated the impact that the curving of an FPA will have on the band gap of bulk InAsSb absorber layers\cite{2021KyrtsosPRAPP}. A natural extension of this investigation would be to study the properties of additional IR materials under strained conditions. 

Over the last few decades, type-II superlattices (T2SLs) have been investigated as a promising material system for use in infrared (IR) devices\,\cite{1984OsbournJVSTB,1987SmithJAP,2011TingSS,2021KwanIPT}. Due to the broken gap alignment of the constituent layers, T2SLs can be grown with arbitrarily small band gaps by adjusting the layer thicknesses. This structural control over the band gap can be achieved while remaining lattice matched to the chosen substrate. This allows for a diverse range of band structures to be achieved, which include various barrier architectures\,\cite{2014Martyniuk,2017DelmasSLMS,2022KlipsteinAPL,2016OlsonAPL,2008NguyenAPL,2009Delaunay,2009NguyenAPL,2012SalihogluAPL,2014Salihoglu,2015HostutSLMS} as well as quantum cascade detectors\,\cite{2013HinkeyJAP,2013Pusz,2022GawronEDL,2022MartyniukPRAPP}. Ga-free (InAs/InAsSb) T2SLs have demonstrated relatively long minority carrier lifetimes in both the mid-wave (MWIR)\,\cite{2013HoglundAPL} and long-wave (LWIR)\,\cite{2011SteenbergenAPL} regimes, which has lead to increased interest in this material system. Despite the long minority carrier lifetimes, n-type Ga-free T2SLs can exhibit poor carrier collection, particularly for longer cut-off wavelengths\,\cite{2020TingMMA}. Ga-free T2SLs grown pseudomorphically on GaSb require thicker periods than the Ga-based variety, due to the smaller band offsets. This combined with structural disorder typically results in poor vertical hole mobility\,\cite{2017OlsonPRA,2020CasiasAPL,2021BellottiPRAPP}. Thus, mobility is a key parameter affecting the performance of n-type Ga-free T2SL-based sensors. While structural disorder can be expected to reduce the vertical carrier mobility\,\cite{2021BellottiPRAPP}, the mutual influence of disorder and mechanically-induced strain on the vertical hole mobility is unclear. A degradation in the vertical hole mobility will reduce carrier collection and ultimately the internal quantum efficiency of the device. Thus, understanding the effect that curving-induced strain and structural disorder have on the transport properties of T2SLs will help predict the impact on device performance, which could ultimately be used as a basis for material-specific design optimization.

The nonequilibrium Green's functions (NEGF) formalism is an optimal methodology for the study of carrier transport in T2SLs. It is a fully quantum mechanical formalism, which means that quantum effects, such as tunneling, are treated implicitly. This is particularly relevant for highly inhomogeneous materials like T2SLs in which one mechanism of transport may not be enough to describe the complex carrier dynamics in these materials\,\cite{1991TsuPRB,1993LaikhtmanPRB}. NEGF calculations can implicitly treat these different mechanisms simultaneously under the same formalism\,\cite{1998WackerPRL,2020BertazziPRAPP}. While NEGF calculations are well suited for treating carrier transport in T2SLs, extracting mobility from them requires careful consideration. A recent paper demonstrated that ballistic effects can be significant in NEGF calculations of T2SLs and for which must be taken into account\,\cite{2023GlennonPRAPP}. When done properly, the NEGF formalism provides an effective method for calculating carrier mobility in T2SLs.

In this work, we perform NEGF calculations of a MWIR InAs/InAsSb SL structure, investigating the impact that structural disorder and external axisymmetric strain in the growth direction has on the vertical hole mobility, and interpreting the implications these effects may have on device performance. The paper is organized as follows: the model for carrier transport as well as the FEA model for describing the mechanical strain are elaborated on in Sec.\,\ref{sec:II}. The calculated vertical hole mobility for varying strain and disorder will be presented in Sec.\,\ref{sec:III}, along with a discussion of the implications these results have for the internal QE of the device. Finally, Sec.\,\ref{sec:IV} will provide a summary of the findings. 
 
\section{\label{sec:II}Methods\protect}
We use the ABAQUS/STANDARD\,\cite{Abaqus} model described in Ref.\,\cite{2021KyrtsosPRAPP} to perform a a finite-element simulation of the strain configuration induced in the curving of an FPA. A full description of the model is provided in the aforementioned paper, but the salient details will be provided here. The mechanical model in the manuscript is for a bulk InAs\textsubscript{0.91}Sb\textsubscript{0.09} absorber\,\cite{2021KyrtsosPRAPP}. While in our case we are modeling a SL, the microscopic scale of the SL structure is not possible to include in the macroscopic die displacement model, so we choose to model the SL as a bulk alloy with an average composition of InAs\textsubscript{0.89}Sb\textsubscript{0.11}. However, as mentioned in Ref.\,\cite{2021KyrtsosPRAPP}, the difference in the strain configuration calculated for InAs\textsubscript{0.82}Sb\textsubscript{0.18} versus InAs\textsubscript{0.91}Sb\textsubscript{0.09} is insignificant. Thus, we use the strain configuration calculated for the InAs\textsubscript{0.91}Sb\textsubscript{0.09} alloy. The die is modeled as a $\mathrm{45\,mm\times 50\,mm}$ rectangular deformable part with a thickness of $\mathrm{100\,\mu m}$. The displacement of the die surface is dictated directly in the simulation, forcing the initially flat die into a spherical segment with a prescribed radius of curvature (ROC) of $\mathrm{70\,mm}$, which is representative of a curved-FPA.

Carrier transport in this study is described using the NEGF formalism, which is a fully quantum-mechanical treatment. A more detailed description of the transport model is presented in Ref.\,\cite{2020BertazziPRAPP}, but the details pertinent to this study will be presented in this section.

The complex band structure of the SL studied in this work was described using the spin-up portion of the 8x8 block diagonalized Hamiltonian within the axial approximation, which incorporates the first conduction band, heavy-hole, light-hole, and spin-orbit split-off bands. The Hamiltonian is solved according to the envelope function approximation in which the envelope functions are expanded in first-order Lagrange polynomials in a finite-element implementation. The band parameters for bulk InAs and InSb were taken from Ref.\,\cite{2012QiaoOEX} and the band gap bowing parameter chosen for the description of bulk InAsSb was taken from Ref.\,\cite{2015WebsterJAP}. The effect of strain on the band structure is incorporated in the Hamiltonian as described in Ref.\,\cite{1995EndersPRB}. 

\begin{figure}
\centering
    \subfloat[\empty\label{subfig:muraki_nd}]{\includegraphics[width=1\columnwidth]{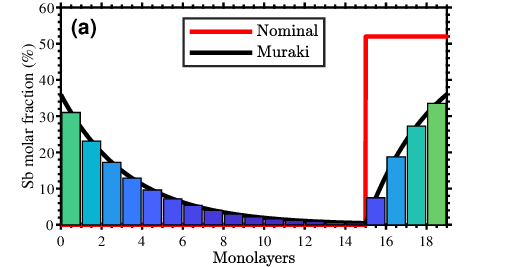}}

    \subfloat[\empty\label{subfig:muraki_fd}]{\includegraphics[width=1\columnwidth]{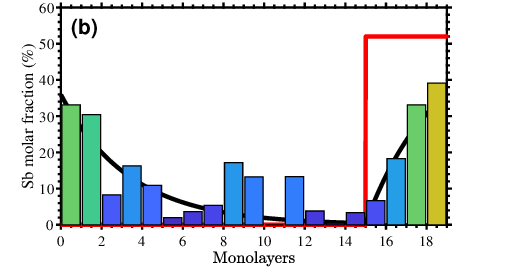}}

    \caption{The Sb concentration profile of an MWIR InAs/InAsSb SL. The nominal (red line) and Muraki model\,\cite{1992MurakiAPL} (black line) concentration profiles are plotted along with the actual profile represented by bar plot with one bar per spatial mesh node in the case of no disorder \protect\subref{subfig:muraki_nd} and disorder \protect\subref{subfig:muraki_fd}. Note, the nominal profile demonstrates the Sb concentration of the Muraki model SL if no Sb segregation occurred. This is a different structure than the ideal SL that is investigated in this study.\label{fig:muraki}}
\end{figure}

To model a realistic SL structure we expand upon the ``compositional-disorder" model presented in Ref\,\cite{2021BellottiPRAPP}, but in this case we incorporate the Muraki Antimony (Sb) segregation model\,\cite{1992MurakiAPL} with the parameters presented in Ref.\,\cite{2018KimJAP}, to model the Sb concentration profile for one SL period. Fig.\,\ref{subfig:muraki_nd} presents the Sb concentration profile for one SL period based on the Muraki model. 
It is demonstrated that a substantial concentration of Sb migrates from the ternary alloy layer to the binary layer, which is consistent with experimentally grown Ga-free SLs\,\cite{2018KimJAP}.

To add disorder to the structure, we select a random number from a normal distribution of Sb concentration with a mean of zero and a selected standard deviation. This value is added to the Sb concentration profile, as defined in Fig.\,\ref{subfig:muraki_nd}, at each node. Fig.\,\ref{subfig:muraki_fd} presents one example of a disordered Muraki Sb concentration profile for one SL period. Multiple differently disordered SL periods can be combined to produce a disordered SL. 

For computational efficiency, transport is restricted to the growth direction of the SL. Acoustic and polar-optical phonon scattering are incorporated via self-energies that are derived according to the self-consistent Born approximation (SCBA) within the deformation potential and Fröhlich formalisms, respectively. Open boundary conditions are incorporated via an additional self-energy term that is calculated according to the complex band structure of the reservoirs\,\cite{2006LuisierPRB}. The Green's functions are solved self-consistently via the Dyson and Keldysh equations\,\cite{2008Aeberhard_PhD,2020BertazziPRAPP}:
\begin{eqnarray}
   &&G^{R}(z_1,z'_1,E)=G_0^{R}(z_1,z'_1,E)\nonumber\\
   &&+\int dz_{2}\int dz_{3}G_0^{R}(z_1,z_2,E)\Sigma^{R}(z_2,z_3,E)G^{R}(z_3,z'_1,E),\nonumber\\
   \label{eq:Dyson}
\end{eqnarray}
\begin{eqnarray}
   &&G^{\lessgtr }(z_1,z'_1,E)=\nonumber\\
   &&\int dz_{2}\int dz_{3}G^{R}(z_1,z_2,E)\Sigma^{\lessgtr }(z_2,z_3,E)G^{A}(z_3,z'_1,E).\nonumber\\
   \label{eq:Keldysh}
\end{eqnarray}

The calculation of Green's functions via the SCBA is computationally prohibitive, thus we solve the Dyson and Keldysh equations using a problem-matched mode-space methodology in which a new basis is chosen from the eigenstates of the noninteracting Hamiltonian. The basis is limited to those functions that fall within an energy range relevant to the carrier conduction in the device while the influence of the other functions is folded into an additional self-energy term to maintain current conservation\,\cite{2006LuisierPRB}. The Dyson and Keldysh equations are solved self-consistently in mode space and then the self-energy is converted to real space in which a final calculation of the Dyson and Keldysh equations are performed to determine the Green's functions. The one-particle properties of the carriers can then be extracted from the Green's functions. The electron (hole) carrier and current densities can be calculated via:\,\cite{2009Steiger_PhD,2008Aeberhard_PhD,2020BertazziPRAPP}
\begin{equation}
   n(z),p(z) =(\mp) \frac{i}{A} \sum_{\bar{k}} \int \frac{dE}{2\pi} G^{\lessgtr }(z,z,\bar{k},E),
   \label{eq:Density}
\end{equation}
\noindent
\begin{equation}
   J_{n,p}(z) = \lim_{z' \to z} \frac{e\hbar}{m_0} \left(\frac{\partial}{\partial z} - \frac{\partial}{\partial z'}\right) \frac{1}{A} \sum_{\bar{k}} \int \frac{dE}{2\pi} G^{\lessgtr }(z,z',\bar{k},E).
   \label{eq:Current}
\end{equation}
The NEGF parameters used in the calculations presented in this study are provided in the Supplementary Material.

Due to the non-negligible ballistic effects in the regime of transport investigated, we use resistance scaling analysis to solve for the vertical hole mobility as described in Ref.\,\cite{2023GlennonPRAPP}:
\begin{equation}
   pR(L_{s}) = pR_c + pR_B + \frac{L_{s}}{\mu e},
   \label{eq:Rn_scale}
\end{equation}
\noindent
where $R_{c}$ and $R_{B}$ are the contact and ballistic resistances, respectively, $L_{s}$ is the simulation size, $\mu$ is the mobility we are extracting, $e$ the elementary charge, and $p$ represents the hole density.

Strain induced by the curving of the FPA can be included in the band structure model via the principle of superposition of linear elasticity in which the curving induced strain is added to the strain induced in the SL due to lattice matching to GaSb. The strain in the active material can have four non-zero components, namely $\vec{S} = \{\epsilon_{xx},\epsilon_{yy},\epsilon_{zz},\gamma_{xy}\}$, which depend upon the physical $x$, $y$, and $z$ coordinates of the die, with the $x$ and $y$ direction being transverse to the SL plane and $z$ being in the growth direction. However, the transport model uses an axial approximation, with an axisymmetric strain, $\epsilon_{biax}$, and a resultant strain in the growth direction given by the following equation:\;\cite{2021KyrtsosPRAPP}:
\begin{equation}
   \epsilon_{zz} = -2\frac{\nu \epsilon_{biax}}{\left(1 - \nu\right)},
   \label{eq:unif_biax}
\end{equation}
\noindent
where $\nu$ is the Poisson's ratio. To obtain a position dependent strain $\epsilon_{biax}$, the strain $\vec{S}$ extracted from the ABAQUS/STANDARD model is transformed to a vector $\vec{S^\prime}$ that adheres to the condition $\epsilon_{biax}$ = $\epsilon_{xx}^\prime$ = $\epsilon_{yy}^\prime$. Using the position dependent biaxial strain vector $\vec{S}_{biax} = \{\epsilon_{biax}, \epsilon_{biax},-2\nu\epsilon_{biax}/(1-\nu), 0\}$ allows us to map physical properties extracted from the NEGF calculations to the strain configurations extracted from the FEA calculations. One can then obtain the physical properties of the SL-based active region as a function of the spatial coordinates on the curved-FPA, and subsequently make predictions about device performance. A regression tool with strong predictive capabilities is necessary to minimize the required number of NEGF calculations for an accurate fit to mitigate the computational costs. 

Of note, the requirement of using an axisymmetric strain state leaves a residual strain ${\vec{S}}_{res}\equiv{\vec{S}}^\prime-{\vec{S}}_{biax}=\{0,0,0,\gamma_{xy}\prime\}$. Shear strains are beyond the scope of the present investigation given the intrinsic limitations of the employed models, and require further study to determine their significance. Therefore, the mapping of properties to a curved sensor should be viewed as qualitative, with the uncertainty increasing with the magnitude of the residual strain vector. The inclusion of the residual strain is non-trivial and would require a similarly transformed Hamiltonian that is not restricted to the axial approximation.

Gaussian processes (GPs) algorithm is a popular and versatile machine learning technique used in various applications, including nonparametric Bayesian regression. One of the primary advantages of GP regression is its ability to provide reliable estimates of model uncertainty, which is particularly useful in real-world scenarios where predictions must be accompanied by a measure of their reliability. The GP approach is based on conditioning a prior distribution of functions on the observed data to obtain the posterior distribution. The mean value and standard deviation of the posterior distribution yields the predictions and their uncertainties, respectively. The GP implementation of the PYTHON SCIKIT-LEARN package\,\cite{pedregosa2011scikitlearn} is used with a kernel combining a radial basis function (RBF) and white noise. The RBF kernel is commonly used in GP regression, as it provides a smooth function that can capture complex relationships between input and output variables. The white noise kernel is included to model the measurement noise, which is often present in real-world data. By combining these two kernels, we are able to achieve a good balance between model flexibility and stability.

To assess the impact that disorder and curving induced strain in the active region on the performance of the FPA, we approximate the internal quantum efficiency for an nBn device in which absorption only occurs in the absorber region. We consider the case of backside illumination with no surface reflection. The quantum efficiency is calculated according to the following equation\;\cite{1981ReineSCSM}:
\begin{eqnarray}
   \eta_{BS} &=& \left(\frac{\alpha L_n}{\alpha^2 L_n^2 - 1}\right)\nonumber\\
   &\times& \left[\frac{\alpha L_n - e^{-\alpha W}\sinh{\left(\frac{W}{L_n}\right)}}{\cosh{\left(\frac{W}{L_n}\right)}}-\alpha L_he^{-\alpha W}\right],
   \label{eq:QE_BS}
\end{eqnarray}
\noindent
where $\alpha$ is the absorption coefficient, $W$ is the absorber thickness, and $L_{n}$ is the minority carrier diffusion length. 

\section{\label{sec:III}Results and Discussion \protect}
We probe the impact that curving-induced strain and structural disorder have on the vertical hole mobility in a MWIR InAs/InAsSb SL with layer thicknesses of $\mathrm{4.5\,nm}$ and $\mathrm{1.2\,nm}$, respectively, latticed matched to a GaSb substrate. The composition of the ternary alloy depends on whether we are studying an ideal SL or a disordered SL.

\begin{figure*}
    \subfloat[\empty\label{subfig:SL_ideal_n12}]{\includegraphics{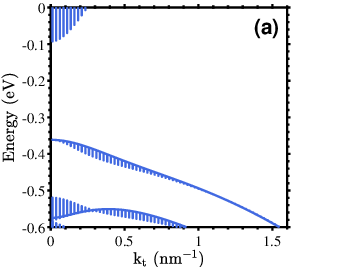}}
    \subfloat[\empty\label{subfig:SL_ideal_n05}]{\includegraphics{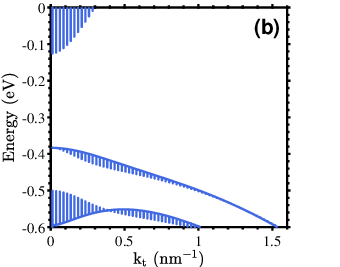}}
    \subfloat[\empty\label{subfig:SL_ideal_n00}]{\includegraphics{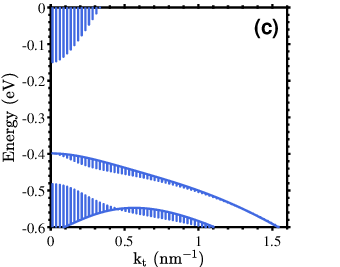}}
    
    \subfloat[\empty\label{subfig:SL_ideal_p05}]{\includegraphics{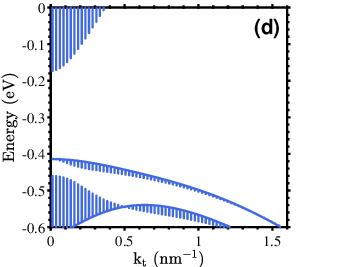}}
    \subfloat[\empty\label{subfig:SL_ideal_p08}]{\includegraphics{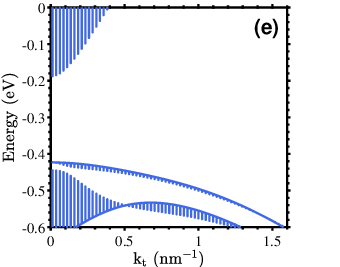}}
    \subfloat[\empty\label{subfig:SL_ideal_p16}]{\includegraphics{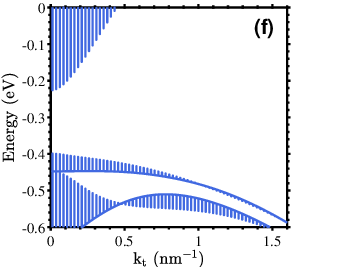}}
    
    \caption{The energy band structure of the ideal SL calculated as a function of the transverse wave vector for external axisymmetric strains of $\mathrm{-1.2\%}$ \protect\subref{subfig:SL_ideal_n12}, $\mathrm{-0.5\%}$ \protect\subref{subfig:SL_ideal_n05}, $\mathrm{0.0\%}$ \protect\subref{subfig:SL_ideal_n00}, $\mathrm{+0.5\%}$ \protect
    \subref{subfig:SL_ideal_p05}, $\mathrm{+0.8\%}$ \protect
    \subref{subfig:SL_ideal_p08}, and $\mathrm{+1.6\%}$ \protect\subref{subfig:SL_ideal_p16}. It is modified with vertical lines representing the dispersion in the vertical wave vector.\label{fig:ideal_bands}}
\end{figure*}

For the ideal structure, we chose the ternary alloy to be InAs\textsubscript{0.65}Sb\textsubscript{0.35} which results in a cutoff wavelength of approximately $\mathrm{5\,\mu m}$. We report the electronic structure of this SL, as calculated using the 4x4\:$k\cdot p$ model with Bloch boundary conditions, in Fig.\,\ref{subfig:SL_ideal_n00}. 
All plots in this manuscript have been generated in MATLAB\,\cite{MATLAB:R2020b_u5} and have been standardized using the toolbox in Ref.\,\cite{atharva2021}. The dispersion in the transverse direction is plotted with branching vertical lines that represent the dispersion in the growth direction for the corresponding in-plane wave number $k_{t}$. Thus, the vertical lines provide an estimate for the miniband width. The top hole miniband is narrow which suggests that the vertical hole mobility may be low even for the ideal case\,\cite{2016TingAPL}.

\begin{figure*}
\centering
    \subfloat[\empty\label{subfig:bgs_dis1}]{\includegraphics{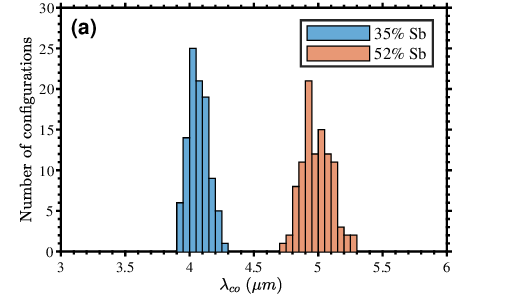}}
    \subfloat[\empty\label{subfig:bgs_dis2}]{\includegraphics{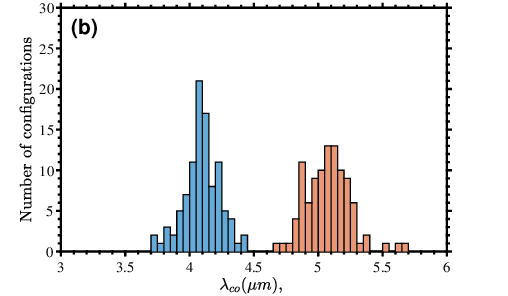}}

    \caption{Histogram of the cutoff wavelengths calculated for low \protect\subref{subfig:bgs_dis1} and high \protect\subref{subfig:bgs_dis2} disorder SLs with ternary alloy layer compositions of InAs\textsubscript{0.65}Sb\textsubscript{0.35} and InAs\textsubscript{0.48}Sb\textsubscript{0.52}.\label{fig:bgs}}
\end{figure*}

In the case of the disordered SLs, the definition of the ternary alloy composition requires further explanation. We investigate three different Muraki model based structures in which the Sb disorder is selected from a normal distribution with standard deviations of 0\% (no disorder), 2\% and 4\%, respectively, resulting in three different magnitudes of disorder. Throughout the rest of this work we will refer to these three structures as ``no'', ``low'', and ``high'' disorder SLs. We choose InAs\textsubscript{0.48}Sb\textsubscript{0.52} as the ``desired'' ternary alloy composition of the disordered structures. The reason for the increase in the Sb composition compared with the ideal SL is that, as shown in Fig.\,\ref{fig:muraki}, the actual Sb composition of the ternary layer will be lower as Sb migrates into the binary layer during growth. Thus, there is a reduction in the band offsets between the peak and trough of the SL band structure for both the VB and CB. A higher Sb concentration is required to compensate for this effect to produce the same band gap. In Fig.\,\ref{fig:bgs} we compute the energy gaps of 100 random unit cells with compositional disorder, each of which is generated by the periodic repetition of the unique SL period, and demonstrate that the distribution of subsequent cutoff wavelengths is centered around approximately $\mathrm{5\,\mu m}$ for both the low and high disorder cases for the InAs\textsubscript{0.48}Sb\textsubscript{0.52} lattice. 
It is also demonstrated that a choice of InAs\textsubscript{0.65}Sb\textsubscript{0.35}, as in the ideal lattice, using the Muraki model and including disorder results in larger band gaps, which is evidence of the reduced band structure offsets between the constituent``layers".

\begin{figure*}
\centering
    \subfloat[\empty\label{subfig:SL_ideal}]{\includegraphics{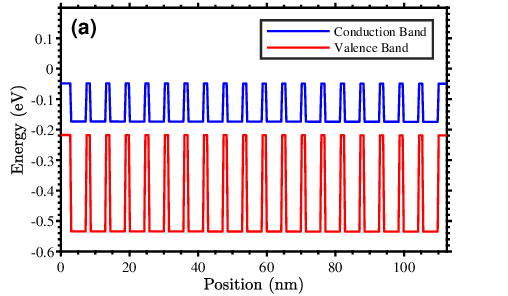}}
    \subfloat[\empty\label{subfig:SL_dis_0}]{\includegraphics{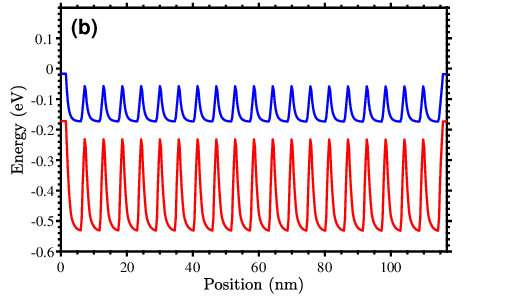}}
    
    \subfloat[\empty\label{subfig:SL_dis_1}]{\includegraphics{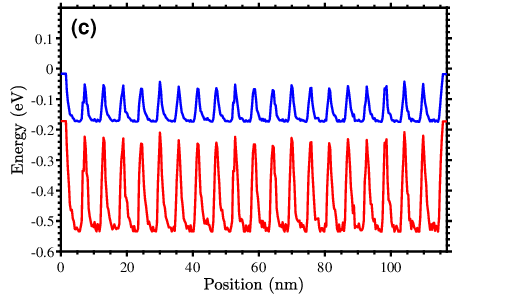}}  
    \subfloat[\empty\label{subfig:SL_dis_2}]{\includegraphics{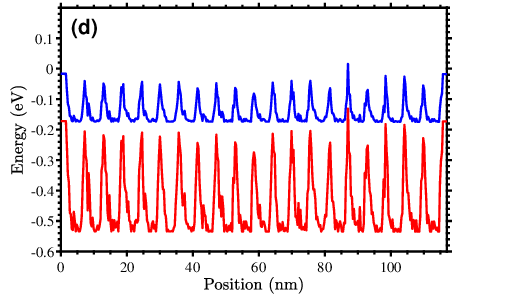}}

    \caption{Band diagrams of the ideal\:\protect\subref{subfig:SL_ideal}, no disorder\:\protect\subref{subfig:SL_dis_0}, low disorder\:\protect\subref{subfig:SL_dis_1}, and high disorder\:\protect\subref{subfig:SL_dis_2} SLs for the simulations with domain size of approximately $\mathrm{120\, nm}$.\label{fig:SL}}
\end{figure*}

We simulate carrier transport using NEGF calculations on finite-sized SL models terminated on each side by open boundary conditions, which act as semi-infinite contact regions. To allow for unimpeded hole transport on both sides of the SL we choose the InAsSb layer for the contact material. In the case of the disordered SLs, this amounts to an ideal InAsSb layer that matches the``desired" Sb concentration selected for the Muraki model. We choose an electric field strength of $\mathrm{100 V/cm}$. Examples of the band structure for each SL investigated (ideal, no disorder, low disorder, and high disorder), with a total simulation domain size of approximately $\mathrm{120\,nm}$ each, are presented in Fig.\,\ref{fig:SL}.
The increasing disorder is evident when comparing the no disorder (Fig.\,\ref{subfig:SL_dis_0}), low disorder (Fig.\,\ref{subfig:SL_dis_1}), and high disorder (Fig.\,\ref{subfig:SL_dis_2}) structures. 

The SL structures that we investigate in this study have non-negligible coherent hole transport which requires the use of the resistance scaling method (Eq.\,\ref{eq:Rn_scale})\,\cite{2023GlennonPRAPP}. A linear curve is fitted to the $Rp$ product of simulations with SL sizes of approximately $\mathrm{120\,nm}$ and $\mathrm{240\,nm}$ to extract the vertical hole mobility. In certain cases, a simulation of $\mathrm{360\,nm}$ is performed as well to ensure that linearity of resistance has been achieved.

\begin{figure}
\includegraphics[width=1\columnwidth]{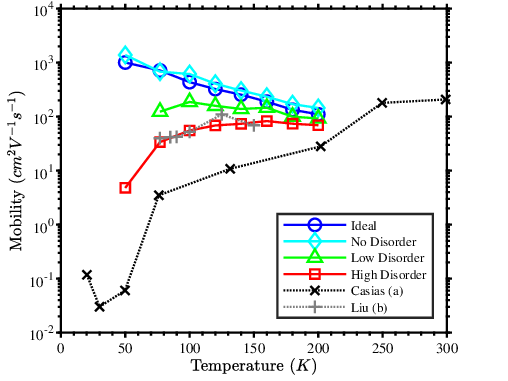}
    \caption{The vertical hole mobility calculated as a function of temperature for the ideal (blue line with circles), no disorder (cyan line with diamonds), low disorder (green line with triangles), and high disorder (red line with squares) SLs. Also plotted are experimental results for vertical hole mobility reported in (a) Ref.\,\cite{2020CasiasAPL} (black dotted line with x's) and (b) Ref.\,\cite{2022LiuAPL} (gray dotted line with +'s) for two different MWIR InAs/InAsSb SLs. The inversion of the temperature dependence of vertical hole mobility at low temperature, exhibited in the experimental data, is reproduced in the disordered structures.\label{fig:exp_results}}
\end{figure}

Before discussing the impact of strain in these structures, it will be beneficial to discuss the effect of disorder in the unstrained case. The vertical hole mobility calculated for the ideal, no disorder, low disorder, and high disorder structures are presented in Fig.\,\ref{fig:exp_results}. 
Additionally, we include experimental vertical hole mobilities reported by Ref.\,\cite{2020CasiasAPL} and Ref.\,\cite{2022LiuAPL} for two different MWIR InAs/InAsSb SLs. The high disorder simulated results are reasonably close in comparison with the experimental data in Ref.\,\cite{2022LiuAPL}. This provides some validity for the selection of the disorder. Furthermore, when observing the mobility for the no disorder structure and comparing it with that of the structures with increasing disorder, it is evident that an inversion of the temperature dependence at low temperature is instigated by the disorder. This behavior is consistent with the temperature trends observed in experimental data for T2SLs\,\cite{2017OlsonPRA,2020CasiasAPL} as well as in the results provided in Ref.\,\cite{2021BellottiPRAPP}, suggesting that holes are propagating through the SL by hopping between localized states\,\cite{2017OlsonPRA,2020BertazziPRAPP}. Finally, the mobility for the no disorder structure is slightly higher than in the ideal structure throughout most of the temperature range. This is likely due to the graded interfaces that result in less hole confinement, demonstrating the impact that Sb segregation alone has on the vertical hole mobility.

\begin{figure*}
\centering
    \subfloat[\empty\label{subfig:SL_ideal_neg}]{\includegraphics{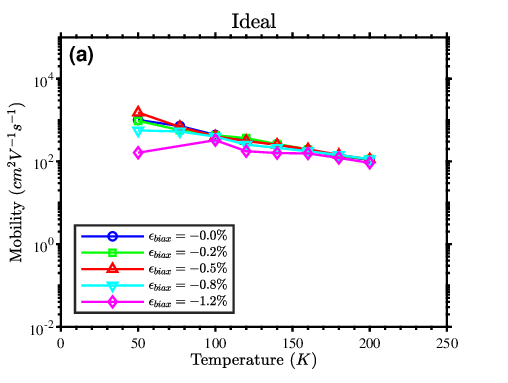}}
    \subfloat[\empty\label{subfig:SL_ideal_pos}]{\includegraphics{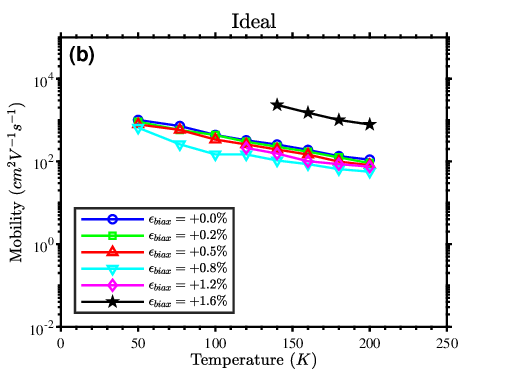}}
    
    \subfloat[\empty\label{subfig:SL_dis_0_neg}]{\includegraphics{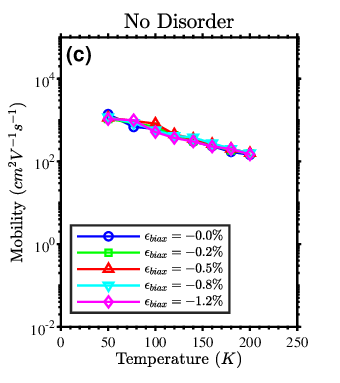}}
    \subfloat[\empty\label{subfig:SL_dis_1_neg}]{\includegraphics{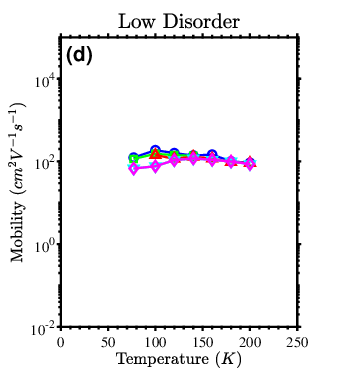}}
    \subfloat[\empty\label{subfig:SL_dis_2_neg}]{\includegraphics{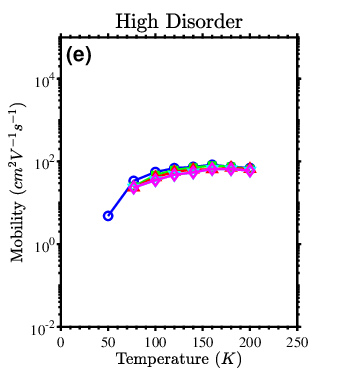}}

    \subfloat[\empty\label{subfig:SL_dis_0_pos}]{\includegraphics{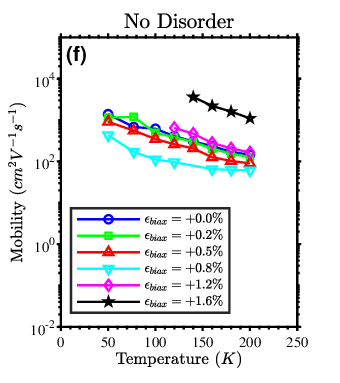}}
    \subfloat[\empty\label{subfig:SL_dis_1_pos}]{\includegraphics{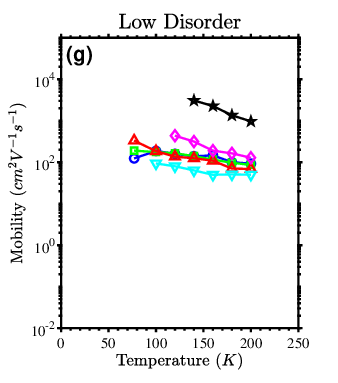}}
    \subfloat[\empty\label{subfig:SL_dis_2_pos}]{\includegraphics{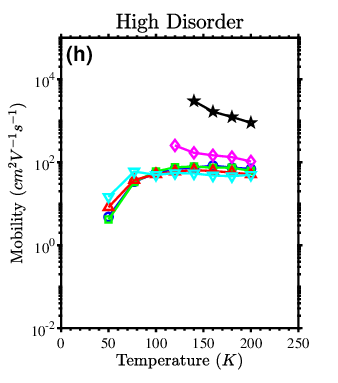}}

    \caption{The vertical hole mobility calculated as a function of temperature for different magnitudes of negative (positive) axisymmetric strain for the ideal\,\protect\subref{subfig:SL_ideal_neg} (\protect\subref{subfig:SL_ideal_pos}), no disorder\,\protect\subref{subfig:SL_dis_0_neg} (\protect\subref{subfig:SL_dis_0_pos}), low disorder\,\protect\subref{subfig:SL_dis_1_neg} (\protect\subref{subfig:SL_dis_1_pos}), and high disorder\,\protect\subref{subfig:SL_dis_2_neg} (\protect\subref{subfig:SL_dis_2_pos}) SLs. The inversion of the temperature dependence of vertical hole mobility at low temperature, exhibited in the experimental data, is reproduced in the disordered structures. The onset of an inversion of the temperature dependence at low temperature occurs with increasing disorder for all strains investigated with exception to the $+1.6\%$ case.\label{fig:mobility}}
\end{figure*}

\begin{figure*}
\centering
    \subfloat[\empty\label{subfig:pos_strain_sweep}]{\includegraphics{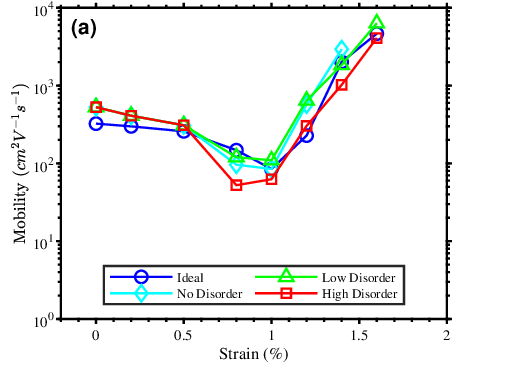}}
    
    \subfloat[\empty\label{subfig:p10_id_bands}]{\includegraphics{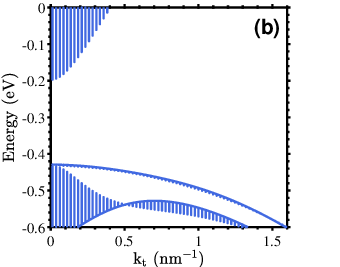}}
    \subfloat[\empty\label{subfig:p10_nD_bands}]{\includegraphics{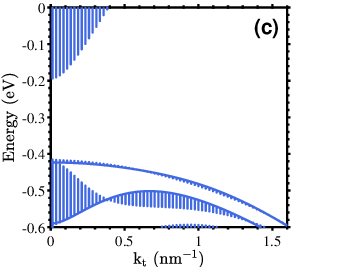}}

    \caption{The vertical hole mobility\:\protect\subref{subfig:120K_mob} calculated as a function of positive axisymmetric strain for each SL structure. Also included are the band structures calculated for the ideal\:\protect\subref{subfig:p10_id_bands} and no disorder\:\protect\subref{subfig:p10_nD_bands} SL in which the HH and LH separation is negligible.\label{fig:pos_strain_sweep}}
\end{figure*}

A set of axisymmetric strains are investigated between $\mathrm{-1.2\%}$ and $\mathrm{+1.6\%}$. Vertical hole mobility calculated as a function of temperature for each strain condition is plotted in Fig.\,\ref{fig:mobility}. 
The disorder-induced inversion of the temperature dependence of the mobility is evident for all external strain values, except for the $\mathrm{+1.2\%}$ and $\mathrm{+1.6\%}$ axisymmetric strain cases. For negative axisymmetric strain, particularly at low temperature, the mobility decreases with increasing strain. The reason for this can be seen in the calculated band structure with applied strain which is presented in Figs.\,\ref{subfig:SL_ideal_n00}, \ref{subfig:SL_ideal_n05}, and \ref{subfig:SL_ideal_n12} for axisymmetric strains of $\mathrm{0.0\%}$, $\mathrm{-0.5\%}$, and $\mathrm{-1.2\%}$, respectively. As negative strain is applied, the approximate HH miniband size becomes narrower, due to reduced dispersion in the transverse direction, suggesting less favorable hole transport properties. On the other hand, the positive axisymmetric case is more complicated. For smaller magnitudes, the mobility decreases. However, at a strain value of $\mathrm{+1.6\%}$ for the ideal and no disorder structures and $\mathrm{+1.2\%}$ for the disordered structures, there is an increase in the mobility over that of the unstrained case. For the $\mathrm{+1.6\%}$ strain case, there is a substantial increase in the hole mobility. This behavior, while more complex than the negative strain case, can also be explained using the calculated band structure of the strained SL (see Figs.\,\ref{subfig:SL_ideal_n00}, \ref{subfig:SL_ideal_p05}, \ref{subfig:SL_ideal_p08}, and \ref{subfig:SL_ideal_p16} for axisymmetric strains of $\mathrm{0.0\%}$, $\mathrm{+0.5\%}$, $\mathrm{+0.8\%}$, and $\mathrm{+1.6\%}$, respectively). For smaller positive axisymmetric strain, the energy offset between the HH and LH bands gets smaller, and the HH and LH bands repel each other. This causes the HH effective mass to increase in the transverse direction. The large increase in hole mobility at $\mathrm{+1.6\%}$ can be attributed to an inversion of the HH and LH character near the gamma point due to the axisymmetric strain reduction in the energy gap between these states. The much lower effective mass of the LH band results in more favorable hole transport properties. To probe this behavior further, we calculated the vertical hole mobility for each SL type at $\mathrm{120\:K}$ with axisymmetric strains of $\mathrm{+1.0\%}$, and $\mathrm{+1.4\%}$. We plot the results in Fig.\:\ref{subfig:pos_strain_sweep} as a function of positive axisymmetric strain.
There is a sharp increase in the mobility above a strain of $\mathrm{+1.0\%}$. Furthermore, we include the calculated band structure of the ideal and no disorder SLs under $\mathrm{+1.0\%}$ axisymmetric strain in Figs.\:\ref{subfig:p10_id_bands} and \ref{subfig:p10_nD_bands}, respectively. It is demonstrated that at this strain the separation between the HH and LH has nearly vanished, indicating that there will be an inversion of the HH and LH at higher strains.

\begin{figure*}
\centering
    \subfloat[\empty\label{subfig:biax_strain}]{\includegraphics{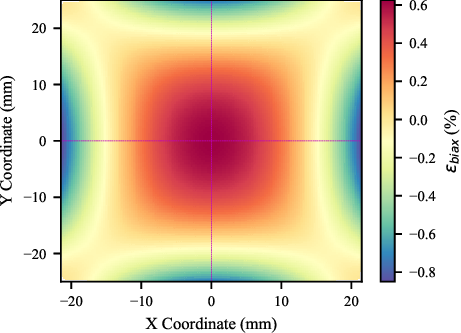}}
    \qquad
    \subfloat[\empty\label{subfig:shear_strain}]{\includegraphics{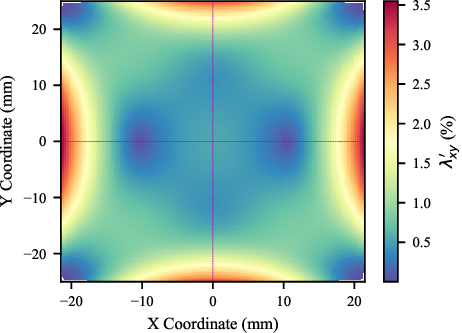}}

    \caption{The projected axisymmetric strain calculated from the strain configuration extracted from the FEA calculations of a curved-FPA die with a ROC of $\mathrm{70\,mm}$\:\protect\subref{subfig:biax_strain}. Also included is the residual strain component from the strainformed strain vector that is not included in the Hamiltonian\:\protect\subref{subfig:shear_strain}.\label{fig:strain}}
\end{figure*}

Fig.\,\ref{subfig:biax_strain} shows the projected axisymmetric strain (Eq.\,(\ref{eq:unif_biax})) colormap for the curved-FPA based on the strain configuration $\{\epsilon_{xx},\epsilon_{yy},\epsilon_{zz},\gamma_{xy}\}$ that was extracted from the FEA calculations.
These strain components are presented in Fig.\,1 of Ref.\,\cite{2021KyrtsosPRAPP}. Also, in Fig.\,\ref{subfig:shear_strain} we present the residual strain component, as described in Sec.\;\ref{sec:II}, which is excluded from the following analysis. To qualitatively probe the performance of the curved-FPA at $120\,K$, we approximate the QE of a backside illuminated nBn device from the calculated vertical hole mobility for each strain and SL structure. The calculated mobility along with a minority carrier lifetime $\tau$ of $\mathrm{100\,ns}$\,\cite{2010DonetskyAPL} are used to estimate the minority carrier diffusion length $L_A$ according to the Einstein relation. We choose an absorption coefficient of $\mathrm{1,000\,cm^{-1}}$ which was experimentally determined for a MWIR Ga-free SL\cite{2019DewamesSPIE}. Then, we use Eq.\,(\ref{eq:QE_BS}) to estimate the QE for nBn devices with an absorber layer thickness $W$ of $\mathrm{4\:\mu m}$. This data set is used to train GP models that can predict the QE for arbitrary axisymmetric strains within the range of strains calculated using NEGF data. Finally, the GP predictions allow us to map approximations for the QE for the two different device models as a function of the spatial coordinates of the die.

\begin{figure*}
\centering
    \subfloat[\empty\label{subfig:120K_mob}]{\includegraphics{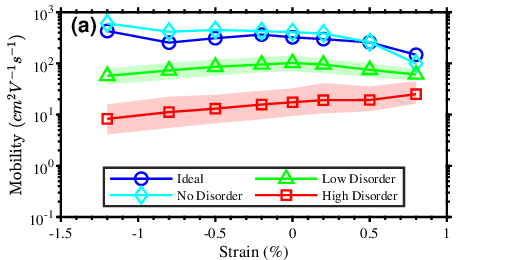}}
    \subfloat[\empty\label{subfig:120K_len}]{\includegraphics{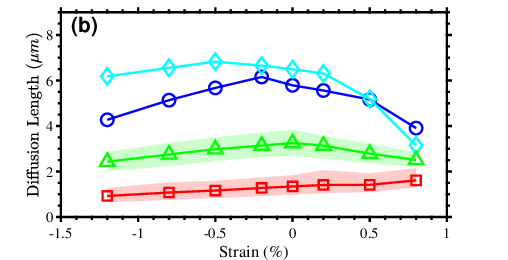}}

    \subfloat[\empty\label{subfig:120K_QE}]{\includegraphics{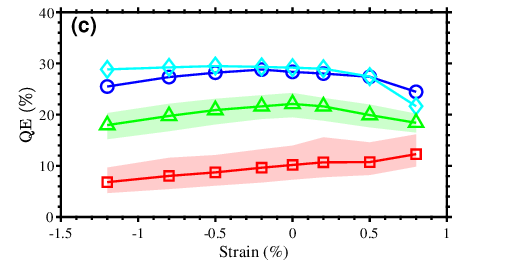}}

    \caption{The vertical hole mobility\:\protect\subref{subfig:120K_mob}, minority hole diffusion length\:\protect\subref{subfig:120K_len}, and the subsequent QE\:\protect\subref{subfig:120K_QE} calculated for a backside illuminated nBn device as a function of axisymmetric strain for the ideal (blue line with circles), no disorder (cyan line with diamonds), low disorder (green line with triangles), and high disorder (red line with squares) SLs. The shaded areas indicate the minimum and maximum values obtained from the different disordered structures for each strain value. The ideal structures do not have disorder, so there is no shaded area.\label{fig:mob_abs_DS}}
\end{figure*}

For the QE estimate we use more comprehensive mobility data sets for low and high magnitudes of disorder. For each magnitude of disorder, we calculate the vertical hole mobility at $\mathrm{120\,K}$ for 10 differently disordered structures for strain configurations of $\mathrm{-1.2\%}$, $\mathrm{-0.8\%}$, $\mathrm{-0.5\%}$, $\mathrm{-0.2\%}$, $\mathrm{0.0\%}$, $\mathrm{+0.2\%}$, $\mathrm{+0.5\%}$, and $\mathrm{+0.8\%}$. Then, we compute the mean values for each strain configuration which are used to train the GP model. Taking the mean of the results over different realizations of the disorder better represents the transport behavior of a structure with a given magnitude of random disorder. The vertical hole mobility, minority hole diffusion length, and QE calculated over different realizations of low and high disorder are presented in Fig.\,\ref{fig:mob_abs_DS} along with the values calculated for the ideal and no disorder SL. 
As previously demonstrated, the disorder negatively impacts the vertical hole mobility throughout the range of axisymmetric strain. The range of values calculated for different realizations of the disorder, measured as a percentage of the average value, increases with increasing disorder. Upon calculation of the QE, it is demonstrated that the estimated QE of the no disorder SL is nearly the same compared to the ideal case. This is likely due to the fact that the diffusion length is longer than the absorber length ($\mathrm{4\:\mu}$) throughout most of the range of strains. Alternatively, the low and high disorder SLs show decreasing QE with increasing disorder as can be expected from the relatively short diffusion lengths.

\begin{figure*}
\centering
    \subfloat[\empty\label{subfig:QE_all}]{\includegraphics{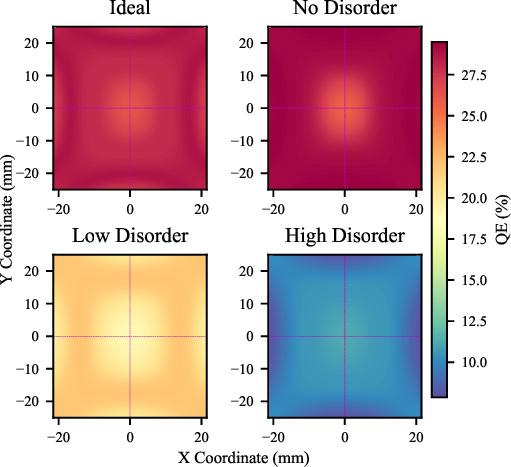}}

    \caption{Approximate internal QE calculated as a function of spatial coordinates of the curved die at $120\,K$ for the ideal, no disorder, low disorder, and high disorder SLs for an nBn detector.\label{fig:QE}}
\end{figure*}

\begin{figure*}
\centering
    \subfloat[\empty\label{subfig:QE_id_2s}]{\includegraphics{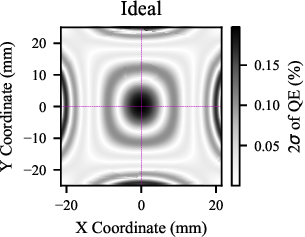}}
    \qquad
    \subfloat[\empty\label{subfig:QE_nd_2s}]{\includegraphics{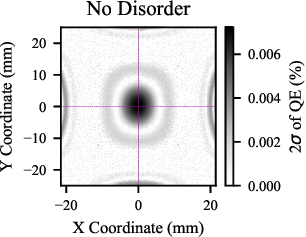}}
    
    \subfloat[\empty\label{subfig:QE_ld_2s}]{\includegraphics{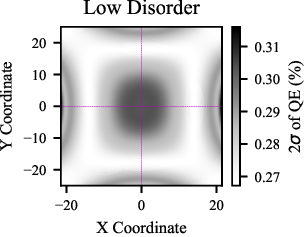}}
    \qquad
    \subfloat[\empty\label{subfig:QE_hd_2s}]{\includegraphics{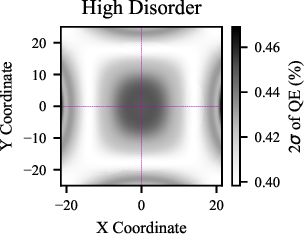}}
    
    \caption{$2\sigma$ of approximated QE calculated as a function of spatial coordinates of the curved die at $\mathrm{120\,K}$ for the ideal, no disorder, low disorder, and high disorder SLs.\label{fig:QE_2s}}
\end{figure*}

Fig.\,\ref{fig:QE} presents the estimated QE for a curved-FPA with ideal, no disorder, low disorder, and high disorder structures.
What is immediately apparent is that the impact on the QE from the strain is relatively small throughout the majority of the FPA in comparison to the impact of disorder. For all four SL types the majority of the FPA exhibits internal QE values within a range of approximately 1\% nominal QE. However, throughout the high disorder device the internal QE values are between 14\% to 22\% nominal QE value lower than those demonstrated in the no disorder case. Thus, it can be expected that curving the FPA to an ROC of $\mathrm{70\,mm}$ is unlikely to dramatically impact the QE with all other properties being equal. On the other hand, the performance can be improved significantly through optimizing the growth process to achieve SLs with reduced disorder. Plots demonstrating the $2\sigma$ uncertainty associated with each device in Fig.\;\ref{fig:QE}, providing the 95\% confidence interval of our results, are presented in Fig.\;\ref{fig:QE_2s}.

\begin{figure*}
\centering
    \subfloat[\empty\label{subfig:QE_pos_biax_id}]{\includegraphics{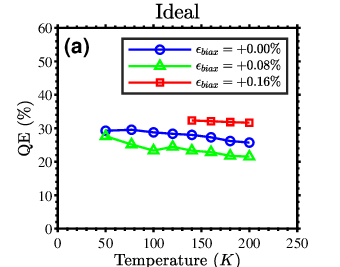}}
    \subfloat[\empty\label{subfig:QE_pos_biax_nD}]{\includegraphics{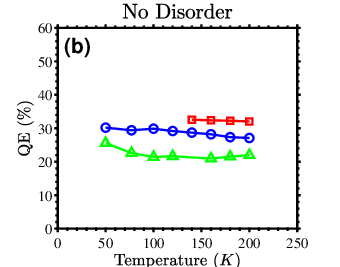}}
    
    \subfloat[\empty\label{subfig:QE_pos_biax_hD}]{\includegraphics{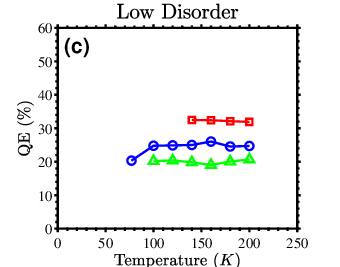}}
    \subfloat[\empty\label{subfig:QE_pos_biax_fD}]{\includegraphics{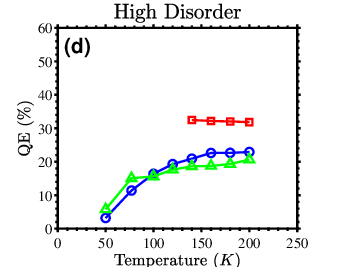}}

    \caption{The internal QE estimated as a function of temperature for the ideal\,\protect\subref{subfig:QE_pos_biax_id}, no disorder\,\protect\subref{subfig:QE_pos_biax_nD}, low disorder\,\protect\subref{subfig:QE_pos_biax_hD}, and high disorder\,\protect\subref{subfig:QE_pos_biax_fD} SLs under $\mathrm{+0.0\%}$ (blue line with circles), $\mathrm{+0.8\%}$ (green line with triangles), and $\mathrm{+1.6\%}$ (red line with squares) axisymmetric strain.\label{fig:QE_pos_biax}}
\end{figure*}

Our NEGF calculations suggest that it may be possible to further improve the performance of a curved-FPA by inducing a higher magnitude of positive axisymmetric strain. This is demonstrated in Fig.\,\ref{fig:QE_pos_biax} in which we plot the quantum efficiency as a function of temperature using Eq.\,(\ref{eq:QE_BS}) for axisymmetric strains of $\mathrm{0.0\%}$, $\mathrm{+0.8\%}$, and $\mathrm{+1.6\%}$. 
There is an improvement in the QE to approximately $34\%$ for all SLs structures at $\mathrm{140\,K}$ and $\mathrm{+1.6\%}$ axisymmetric strain. The increase in QE for the $\mathrm{+1.6\%}$ case is a result of the considerable increase in vertical hole mobility. The die we are investigating, with an ROC of $\mathrm{70\,mm}$, exhibits a maximum projected axisymmetric strain of just over $\mathrm{+0.6\%}$. Fig.\,\ref{fig:QE_pos_biax} suggests that if greater tensile strain can be mechanically tolerated, then this effect may be used to overcome the degradation in QE due to SL disorder and improved device performance could be realized.

\section{\label{sec:IV}Conclusion \protect}
In conclusion, a study of the internal QE predicted for a MWIR Ga-free T2SL-based curved-FPA is presented which demonstrates a framework for studying the transport properties of externally strained T2SLs based on a combination of FEA modeling, NEGF calculations, and machine learning. The strain configuration in the active region of the FPA induced by the curving procedure is modeled using FEA and is projected onto the growth direction axisymmetric strain axes. Then, a rigorous one-dimensional transport model based on the $k\cdot p$ band structure model and NEGF is used to calculate the vertical hole mobility as a function of temperature for differently disordered SLs subject to a set of axisymmetric strain conditions. Then, the internal QE is approximated for an nBn photodetector. Finally, a GP model is trained to map the predicted QE onto the curved-FPA model using the strain configuration that was predicted in FEA, reducing the number of required NEGF calculations and thus the computational cost of the study. The final result is the predicted internal QE across the spatial coordinates of the FPA for different magnitudes of SL disorder. Additionally, it is shown that it may be possible to enhance the QE by implementing large positive axisymmetric strain. This procedure can be used more generally to investigate the transport properties of externally strained materials and predict the impact on device performance. 

\begin{acknowledgments}
The work was supported by the Defense Advanced Research Projects Agency (DARPA) Focal Arrays for Curved Infrared Imagers (FOCII) program (Contract No. N66001-20-C-4011), managed initially by Dr.\:Whitney\:Mason and then subsequently by Trish Veeder. The NEGF code was developed by Prof. Francesco Bertazzi and Prof. Alberto Tibaldi at Politecnico di Torino in Turin, Italy.

The views, opinions and/or findings expressed are those of the author and should not be interpreted as representing the official views or policies of the Department of Defense or the U.S. Government.

\end{acknowledgments}

\bibliography{riviabbr,paper,web,conference,textbook,misc}

\begin{thebibliography}{51}%
\makeatletter
\providecommand \@ifxundefined [1]{%
 \@ifx{#1\undefined}
}%
\providecommand \@ifnum [1]{%
 \ifnum #1\expandafter \@firstoftwo
 \else \expandafter \@secondoftwo
 \fi
}%
\providecommand \@ifx [1]{%
 \ifx #1\expandafter \@firstoftwo
 \else \expandafter \@secondoftwo
 \fi
}%
\providecommand \natexlab [1]{#1}%
\providecommand \enquote  [1]{``#1''}%
\providecommand \bibnamefont  [1]{#1}%
\providecommand \bibfnamefont [1]{#1}%
\providecommand \citenamefont [1]{#1}%
\providecommand \href@noop [0]{\@secondoftwo}%
\providecommand \href [0]{\begingroup \@sanitize@url \@href}%
\providecommand \@href[1]{\@@startlink{#1}\@@href}%
\providecommand \@@href[1]{\endgroup#1\@@endlink}%
\providecommand \@sanitize@url [0]{\catcode `\\12\catcode `\$12\catcode `\&12\catcode `\#12\catcode `\^12\catcode `\_12\catcode `\%12\relax}%
\providecommand \@@startlink[1]{}%
\providecommand \@@endlink[0]{}%
\providecommand \url  [0]{\begingroup\@sanitize@url \@url }%
\providecommand \@url [1]{\endgroup\@href {#1}{\urlprefix }}%
\providecommand \urlprefix  [0]{URL }%
\providecommand \Eprint [0]{\href }%
\providecommand \doibase [0]{https://doi.org/}%
\providecommand \selectlanguage [0]{\@gobble}%
\providecommand \bibinfo  [0]{\@secondoftwo}%
\providecommand \bibfield  [0]{\@secondoftwo}%
\providecommand \translation [1]{[#1]}%
\providecommand \BibitemOpen [0]{}%
\providecommand \bibitemStop [0]{}%
\providecommand \bibitemNoStop [0]{.\EOS\space}%
\providecommand \EOS [0]{\spacefactor3000\relax}%
\providecommand \BibitemShut  [1]{\csname bibitem#1\endcsname}%
\let\auto@bib@innerbib\@empty
\bibitem [{\citenamefont {Rim}\ \emph {et~al.}(2008)\citenamefont {Rim}, \citenamefont {Catrysse}, \citenamefont {Dinyari}, \citenamefont {Huang},\ and\ \citenamefont {Peumans}}]{2008RimOPG}%
  \BibitemOpen
  \bibfield  {author} {\bibinfo {author} {\bibfnamefont {S.-B.}\ \bibnamefont {Rim}}, \bibinfo {author} {\bibfnamefont {P.~B.}\ \bibnamefont {Catrysse}}, \bibinfo {author} {\bibfnamefont {R.}~\bibnamefont {Dinyari}}, \bibinfo {author} {\bibfnamefont {K.}~\bibnamefont {Huang}},\ and\ \bibinfo {author} {\bibfnamefont {P.}~\bibnamefont {Peumans}},\ }\bibfield  {title} {\bibinfo {title} {The optical advantages of curved focal plane arrays},\ }\href {https://doi.org/10.1364/OE.16.004965} {\bibfield  {journal} {\bibinfo  {journal} {Opt. Express}\ }\textbf {\bibinfo {volume} {16}},\ \bibinfo {pages} {4965} (\bibinfo {year} {2008})}\BibitemShut {NoStop}%
\bibitem [{\citenamefont {Rogalski}\ \emph {et~al.}(2016)\citenamefont {Rogalski}, \citenamefont {Martyniuk},\ and\ \citenamefont {Kopytko}}]{2016RogalskiIOP}%
  \BibitemOpen
  \bibfield  {author} {\bibinfo {author} {\bibfnamefont {A.}~\bibnamefont {Rogalski}}, \bibinfo {author} {\bibfnamefont {P.}~\bibnamefont {Martyniuk}},\ and\ \bibinfo {author} {\bibfnamefont {M.}~\bibnamefont {Kopytko}},\ }\bibfield  {title} {\bibinfo {title} {Challenges of small-pixel infrared detectors: a review},\ }\href {https://doi.org/10.1088/0034-4885/79/4/046501} {\bibfield  {journal} {\bibinfo  {journal} {Reports on Progress in Physics}\ }\textbf {\bibinfo {volume} {79}},\ \bibinfo {pages} {046501} (\bibinfo {year} {2016})}\BibitemShut {NoStop}%
\bibitem [{\citenamefont {Guenter}\ \emph {et~al.}(2017)\citenamefont {Guenter}, \citenamefont {Joshi}, \citenamefont {Stoakley}, \citenamefont {Keefe}, \citenamefont {Geary}, \citenamefont {Freeman}, \citenamefont {Hundley}, \citenamefont {Patterson}, \citenamefont {Hammon}, \citenamefont {Herrera}, \citenamefont {Sherman}, \citenamefont {Nowak}, \citenamefont {Schubert}, \citenamefont {Brewer}, \citenamefont {Yang}, \citenamefont {Mott},\ and\ \citenamefont {McKnight}}]{2017GuenterOPG}%
  \BibitemOpen
  \bibfield  {author} {\bibinfo {author} {\bibfnamefont {B.}~\bibnamefont {Guenter}}, \bibinfo {author} {\bibfnamefont {N.}~\bibnamefont {Joshi}}, \bibinfo {author} {\bibfnamefont {R.}~\bibnamefont {Stoakley}}, \bibinfo {author} {\bibfnamefont {A.}~\bibnamefont {Keefe}}, \bibinfo {author} {\bibfnamefont {K.}~\bibnamefont {Geary}}, \bibinfo {author} {\bibfnamefont {R.}~\bibnamefont {Freeman}}, \bibinfo {author} {\bibfnamefont {J.}~\bibnamefont {Hundley}}, \bibinfo {author} {\bibfnamefont {P.}~\bibnamefont {Patterson}}, \bibinfo {author} {\bibfnamefont {D.}~\bibnamefont {Hammon}}, \bibinfo {author} {\bibfnamefont {G.}~\bibnamefont {Herrera}}, \bibinfo {author} {\bibfnamefont {E.}~\bibnamefont {Sherman}}, \bibinfo {author} {\bibfnamefont {A.}~\bibnamefont {Nowak}}, \bibinfo {author} {\bibfnamefont {R.}~\bibnamefont {Schubert}}, \bibinfo {author} {\bibfnamefont {P.}~\bibnamefont {Brewer}}, \bibinfo {author} {\bibfnamefont {L.}~\bibnamefont {Yang}}, \bibinfo {author} {\bibfnamefont {R.}~\bibnamefont {Mott}},\ and\
  \bibinfo {author} {\bibfnamefont {G.}~\bibnamefont {McKnight}},\ }\bibfield  {title} {\bibinfo {title} {Highly curved image sensors: a practical approach for improved optical performance},\ }\href {https://doi.org/10.1364/OE.25.013010} {\bibfield  {journal} {\bibinfo  {journal} {Opt. Express}\ }\textbf {\bibinfo {volume} {25}},\ \bibinfo {pages} {13010} (\bibinfo {year} {2017})}\BibitemShut {NoStop}%
\bibitem [{\citenamefont {O'Masta}\ \emph {et~al.}(2022)\citenamefont {O'Masta}, \citenamefont {Nguyen}, \citenamefont {Gurga}, \citenamefont {Sasse}, \citenamefont {Hempe}, \citenamefont {Neuhaus}, \citenamefont {Clough}, \citenamefont {Hundley}, \citenamefont {Patterson}, \citenamefont {Jenkins}, \citenamefont {Chen}, \citenamefont {Jacques}, \citenamefont {Linton}, \citenamefont {Perez}, \citenamefont {Ysseldyk}, \citenamefont {Wang}, \citenamefont {Tang}, \citenamefont {Niwa}, \citenamefont {Schaedler}, \citenamefont {Kyrtsos}, \citenamefont {Glennon}, \citenamefont {Glasmann}, \citenamefont {Bellotti},\ and\ \citenamefont {McKnight}}]{2022OMastaITA}%
  \BibitemOpen
  \bibfield  {author} {\bibinfo {author} {\bibfnamefont {M.~R.}\ \bibnamefont {O'Masta}}, \bibinfo {author} {\bibfnamefont {B.-M.}\ \bibnamefont {Nguyen}}, \bibinfo {author} {\bibfnamefont {A.}~\bibnamefont {Gurga}}, \bibinfo {author} {\bibfnamefont {T.}~\bibnamefont {Sasse}}, \bibinfo {author} {\bibfnamefont {B.}~\bibnamefont {Hempe}}, \bibinfo {author} {\bibfnamefont {C.}~\bibnamefont {Neuhaus}}, \bibinfo {author} {\bibfnamefont {E.~C.}\ \bibnamefont {Clough}}, \bibinfo {author} {\bibfnamefont {J.}~\bibnamefont {Hundley}}, \bibinfo {author} {\bibfnamefont {P.~R.}\ \bibnamefont {Patterson}}, \bibinfo {author} {\bibfnamefont {J.}~\bibnamefont {Jenkins}}, \bibinfo {author} {\bibfnamefont {M.}~\bibnamefont {Chen}}, \bibinfo {author} {\bibfnamefont {G.}~\bibnamefont {Jacques}}, \bibinfo {author} {\bibfnamefont {S.}~\bibnamefont {Linton}}, \bibinfo {author} {\bibfnamefont {F.}~\bibnamefont {Perez}}, \bibinfo {author} {\bibfnamefont {C.~V.}\ \bibnamefont {Ysseldyk}}, \bibinfo {author} {\bibfnamefont
  {E.}~\bibnamefont {Wang}}, \bibinfo {author} {\bibfnamefont {Y.}~\bibnamefont {Tang}}, \bibinfo {author} {\bibfnamefont {K.}~\bibnamefont {Niwa}}, \bibinfo {author} {\bibfnamefont {T.}~\bibnamefont {Schaedler}}, \bibinfo {author} {\bibfnamefont {A.}~\bibnamefont {Kyrtsos}}, \bibinfo {author} {\bibfnamefont {J.}~\bibnamefont {Glennon}}, \bibinfo {author} {\bibfnamefont {A.}~\bibnamefont {Glasmann}}, \bibinfo {author} {\bibfnamefont {E.}~\bibnamefont {Bellotti}},\ and\ \bibinfo {author} {\bibfnamefont {G.}~\bibnamefont {McKnight}},\ }\bibfield  {title} {\bibinfo {title} {{Curving of large-format infrared sensors}},\ }in\ \href {https://doi.org/10.1117/12.2618122} {\emph {\bibinfo {booktitle} {Infrared Technology and Applications XLVIII}}},\ Vol.\ \bibinfo {volume} {12107},\ \bibinfo {editor} {edited by\ \bibinfo {editor} {\bibfnamefont {B.~F.}\ \bibnamefont {Andresen}}, \bibinfo {editor} {\bibfnamefont {G.~F.}\ \bibnamefont {Fulop}},\ and\ \bibinfo {editor} {\bibfnamefont {L.}~\bibnamefont {Zheng}}},\
  \bibinfo {organization} {International Society for Optics and Photonics}\ (\bibinfo  {publisher} {SPIE},\ \bibinfo {year} {2022})\ p.\ \bibinfo {pages} {121071S}\BibitemShut {NoStop}%
\bibitem [{\citenamefont {Kyrtsos}\ \emph {et~al.}(2021)\citenamefont {Kyrtsos}, \citenamefont {Glennon}, \citenamefont {Glasmann}, \citenamefont {O'Masta}, \citenamefont {Nguyen},\ and\ \citenamefont {Bellotti}}]{2021KyrtsosPRAPP}%
  \BibitemOpen
  \bibfield  {author} {\bibinfo {author} {\bibfnamefont {A.}~\bibnamefont {Kyrtsos}}, \bibinfo {author} {\bibfnamefont {J.}~\bibnamefont {Glennon}}, \bibinfo {author} {\bibfnamefont {A.}~\bibnamefont {Glasmann}}, \bibinfo {author} {\bibfnamefont {M.~R.}\ \bibnamefont {O'Masta}}, \bibinfo {author} {\bibfnamefont {B.-M.}\ \bibnamefont {Nguyen}},\ and\ \bibinfo {author} {\bibfnamefont {E.}~\bibnamefont {Bellotti}},\ }\bibfield  {title} {\bibinfo {title} {Machine-learning-assisted first-principles calculations of strained ${\mathrm{inas}}_{1\ensuremath{-}x}{\mathrm{sb}}_{x}$ alloys for curved focal-plane arrays},\ }\href {https://doi.org/10.1103/PhysRevApplied.15.064008} {\bibfield  {journal} {\bibinfo  {journal} {Phys. Rev. Appl.}\ }\textbf {\bibinfo {volume} {15}},\ \bibinfo {pages} {064008} (\bibinfo {year} {2021})}\BibitemShut {NoStop}%
\bibitem [{\citenamefont {Osbourn}(1984)}]{1984OsbournJVSTB}%
  \BibitemOpen
  \bibfield  {author} {\bibinfo {author} {\bibfnamefont {G.~C.}\ \bibnamefont {Osbourn}},\ }\bibfield  {title} {\bibinfo {title} {Inassb strained‐layer superlattices for long wavelength detector applications},\ }\href {https://doi.org/10.1116/1.582772} {\bibfield  {journal} {\bibinfo  {journal} {J. Vac. Sci. Technol. B}\ }\textbf {\bibinfo {volume} {2}},\ \bibinfo {pages} {176} (\bibinfo {year} {1984})}\BibitemShut {NoStop}%
\bibitem [{\citenamefont {Smith}\ and\ \citenamefont {Mailhiot}(1987)}]{1987SmithJAP}%
  \BibitemOpen
  \bibfield  {author} {\bibinfo {author} {\bibfnamefont {D.~L.}\ \bibnamefont {Smith}}\ and\ \bibinfo {author} {\bibfnamefont {C.}~\bibnamefont {Mailhiot}},\ }\bibfield  {title} {\bibinfo {title} {Proposal for strained type ii superlattice infrared detectors},\ }\href {https://doi.org/10.1063/1.339468} {\bibfield  {journal} {\bibinfo  {journal} {J. Appl. Phys.}\ }\textbf {\bibinfo {volume} {62}},\ \bibinfo {pages} {2545} (\bibinfo {year} {1987})},\ \Eprint {https://arxiv.org/abs/https://doi.org/10.1063/1.339468} {https://doi.org/10.1063/1.339468} \BibitemShut {NoStop}%
\bibitem [{\citenamefont {Ting}\ \emph {et~al.}(2011)\citenamefont {Ting}, \citenamefont {Soibel}, \citenamefont {Höglund}, \citenamefont {Nguyen}, \citenamefont {Hill}, \citenamefont {Khoshakhlagh},\ and\ \citenamefont {Gunapala}}]{2011TingSS}%
  \BibitemOpen
  \bibfield  {author} {\bibinfo {author} {\bibfnamefont {D.~Z.-Y.}\ \bibnamefont {Ting}}, \bibinfo {author} {\bibfnamefont {A.}~\bibnamefont {Soibel}}, \bibinfo {author} {\bibfnamefont {L.}~\bibnamefont {Höglund}}, \bibinfo {author} {\bibfnamefont {J.}~\bibnamefont {Nguyen}}, \bibinfo {author} {\bibfnamefont {C.~J.}\ \bibnamefont {Hill}}, \bibinfo {author} {\bibfnamefont {A.}~\bibnamefont {Khoshakhlagh}},\ and\ \bibinfo {author} {\bibfnamefont {S.~D.}\ \bibnamefont {Gunapala}},\ }\bibfield  {title} {\bibinfo {title} {Chapter 1 - type-ii superlattice infrared detectors},\ }in\ \href {https://doi.org/https://doi.org/10.1016/B978-0-12-381337-4.00001-2} {\emph {\bibinfo {booktitle} {Advances in Infrared Photodetectors}}},\ \bibinfo {series} {Semiconductors and Semimetals}, Vol.~\bibinfo {volume} {84},\ \bibinfo {editor} {edited by\ \bibinfo {editor} {\bibfnamefont {S.~D.}\ \bibnamefont {Gunapala}}, \bibinfo {editor} {\bibfnamefont {D.~R.}\ \bibnamefont {Rhiger}},\ and\ \bibinfo {editor} {\bibfnamefont
  {C.}~\bibnamefont {Jagadish}}}\ (\bibinfo  {publisher} {Elsevier},\ \bibinfo {year} {2011})\ pp.\ \bibinfo {pages} {1--57}\BibitemShut {NoStop}%
\bibitem [{\citenamefont {Kwan}\ \emph {et~al.}(2021)\citenamefont {Kwan}, \citenamefont {Kesaria}, \citenamefont {Anyebe},\ and\ \citenamefont {Huffaker}}]{2021KwanIPT}%
  \BibitemOpen
  \bibfield  {author} {\bibinfo {author} {\bibfnamefont {D.}~\bibnamefont {Kwan}}, \bibinfo {author} {\bibfnamefont {M.}~\bibnamefont {Kesaria}}, \bibinfo {author} {\bibfnamefont {E.~A.}\ \bibnamefont {Anyebe}},\ and\ \bibinfo {author} {\bibfnamefont {D.}~\bibnamefont {Huffaker}},\ }\bibfield  {title} {\bibinfo {title} {Recent trends in 8--14 $\mu$m type-ii superlattice infrared detectors},\ }\href {https://www.sciencedirect.com/science/article/pii/S1350449521001286} {\bibfield  {journal} {\bibinfo  {journal} {Infrared Physics {\&} Technology}\ }\textbf {\bibinfo {volume} {116}},\ \bibinfo {pages} {103756} (\bibinfo {year} {2021})}\BibitemShut {NoStop}%
\bibitem [{\citenamefont {Martyniuk}\ \emph {et~al.}(2014)\citenamefont {Martyniuk}, \citenamefont {Kopytko},\ and\ \citenamefont {Rogalski}}]{2014Martyniuk}%
  \BibitemOpen
  \bibfield  {author} {\bibinfo {author} {\bibfnamefont {P.}~\bibnamefont {Martyniuk}}, \bibinfo {author} {\bibfnamefont {M.}~\bibnamefont {Kopytko}},\ and\ \bibinfo {author} {\bibfnamefont {A.}~\bibnamefont {Rogalski}},\ }\bibfield  {title} {\bibinfo {title} {Barrier infrared detectors},\ }\href {https://doi.org/10.2478/s11772-014-0187-x} {\bibfield  {journal} {\bibinfo  {journal} {Opto-Electronics Review}\ }\textbf {\bibinfo {volume} {22}},\ \bibinfo {pages} {127} (\bibinfo {year} {2014})}\BibitemShut {NoStop}%
\bibitem [{\citenamefont {Delmas}\ \emph {et~al.}(2017)\citenamefont {Delmas}, \citenamefont {Rossignol}, \citenamefont {Rodriguez},\ and\ \citenamefont {Christol}}]{2017DelmasSLMS}%
  \BibitemOpen
  \bibfield  {author} {\bibinfo {author} {\bibfnamefont {M.}~\bibnamefont {Delmas}}, \bibinfo {author} {\bibfnamefont {R.}~\bibnamefont {Rossignol}}, \bibinfo {author} {\bibfnamefont {J.}~\bibnamefont {Rodriguez}},\ and\ \bibinfo {author} {\bibfnamefont {P.}~\bibnamefont {Christol}},\ }\bibfield  {title} {\bibinfo {title} {Design of inas/gasb superlattice infrared barrier detectors},\ }\href {https://doi.org/https://doi.org/10.1016/j.spmi.2017.03.001} {\bibfield  {journal} {\bibinfo  {journal} {Superlatt. Microstruct.}\ }\textbf {\bibinfo {volume} {104}},\ \bibinfo {pages} {402} (\bibinfo {year} {2017})}\BibitemShut {NoStop}%
\bibitem [{\citenamefont {Klipstein}(2022)}]{2022KlipsteinAPL}%
  \BibitemOpen
  \bibfield  {author} {\bibinfo {author} {\bibfnamefont {P.~C.}\ \bibnamefont {Klipstein}},\ }\bibfield  {title} {\bibinfo {title} {Perspective on iii–v barrier detectors},\ }\href {https://doi.org/10.1063/5.0084100} {\bibfield  {journal} {\bibinfo  {journal} {Appl. Phys. Lett.}\ }\textbf {\bibinfo {volume} {120}},\ \bibinfo {pages} {060502} (\bibinfo {year} {2022})},\ \Eprint {https://arxiv.org/abs/https://doi.org/10.1063/5.0084100} {https://doi.org/10.1063/5.0084100} \BibitemShut {NoStop}%
\bibitem [{\citenamefont {Olson}\ \emph {et~al.}(2016)\citenamefont {Olson}, \citenamefont {Kim}, \citenamefont {Kadlec}, \citenamefont {Klem}, \citenamefont {Hawkins}, \citenamefont {Coon}, \citenamefont {Fortune}, \citenamefont {Tauke-Pedretti}, \citenamefont {Cavaliere},\ and\ \citenamefont {Shaner}}]{2016OlsonAPL}%
  \BibitemOpen
  \bibfield  {author} {\bibinfo {author} {\bibfnamefont {B.~V.}\ \bibnamefont {Olson}}, \bibinfo {author} {\bibfnamefont {J.~K.}\ \bibnamefont {Kim}}, \bibinfo {author} {\bibfnamefont {E.~A.}\ \bibnamefont {Kadlec}}, \bibinfo {author} {\bibfnamefont {J.~F.}\ \bibnamefont {Klem}}, \bibinfo {author} {\bibfnamefont {S.~D.}\ \bibnamefont {Hawkins}}, \bibinfo {author} {\bibfnamefont {W.~T.}\ \bibnamefont {Coon}}, \bibinfo {author} {\bibfnamefont {T.~R.}\ \bibnamefont {Fortune}}, \bibinfo {author} {\bibfnamefont {A.}~\bibnamefont {Tauke-Pedretti}}, \bibinfo {author} {\bibfnamefont {M.~A.}\ \bibnamefont {Cavaliere}},\ and\ \bibinfo {author} {\bibfnamefont {E.~A.}\ \bibnamefont {Shaner}},\ }\bibfield  {title} {\bibinfo {title} {Optical and electrical properties of narrow-bandgap infrared w-structure superlattices incorporating alas/alsb/alas barrier layers},\ }\href {https://doi.org/10.1063/1.4954649} {\bibfield  {journal} {\bibinfo  {journal} {Appl. Phys. Lett.}\ }\textbf {\bibinfo {volume} {108}},\ \bibinfo {pages}
  {252104} (\bibinfo {year} {2016})},\ \Eprint {https://arxiv.org/abs/https://doi.org/10.1063/1.4954649} {https://doi.org/10.1063/1.4954649} \BibitemShut {NoStop}%
\bibitem [{\citenamefont {Nguyen}\ \emph {et~al.}(2008)\citenamefont {Nguyen}, \citenamefont {Hoffman}, \citenamefont {Delaunay}, \citenamefont {Huang}, \citenamefont {Razeghi},\ and\ \citenamefont {Pellegrino}}]{2008NguyenAPL}%
  \BibitemOpen
  \bibfield  {author} {\bibinfo {author} {\bibfnamefont {B.-M.}\ \bibnamefont {Nguyen}}, \bibinfo {author} {\bibfnamefont {D.}~\bibnamefont {Hoffman}}, \bibinfo {author} {\bibfnamefont {P.-Y.}\ \bibnamefont {Delaunay}}, \bibinfo {author} {\bibfnamefont {E.~K.-W.}\ \bibnamefont {Huang}}, \bibinfo {author} {\bibfnamefont {M.}~\bibnamefont {Razeghi}},\ and\ \bibinfo {author} {\bibfnamefont {J.}~\bibnamefont {Pellegrino}},\ }\bibfield  {title} {\bibinfo {title} {Band edge tunability of m-structure for heterojunction design in sb based type ii superlattice photodiodes},\ }\href {https://doi.org/10.1063/1.3005196} {\bibfield  {journal} {\bibinfo  {journal} {Appl. Phys. Lett.}\ }\textbf {\bibinfo {volume} {93}},\ \bibinfo {pages} {163502} (\bibinfo {year} {2008})},\ \Eprint {https://arxiv.org/abs/https://doi.org/10.1063/1.3005196} {https://doi.org/10.1063/1.3005196} \BibitemShut {NoStop}%
\bibitem [{\citenamefont {Delaunay}\ \emph {et~al.}(2009)\citenamefont {Delaunay}, \citenamefont {Nguyen}, \citenamefont {Hoffman}, \citenamefont {Huang},\ and\ \citenamefont {Razeghi}}]{2009Delaunay}%
  \BibitemOpen
  \bibfield  {author} {\bibinfo {author} {\bibfnamefont {P.-Y.}\ \bibnamefont {Delaunay}}, \bibinfo {author} {\bibfnamefont {B.~M.}\ \bibnamefont {Nguyen}}, \bibinfo {author} {\bibfnamefont {D.}~\bibnamefont {Hoffman}}, \bibinfo {author} {\bibfnamefont {E.~K.-W.}\ \bibnamefont {Huang}},\ and\ \bibinfo {author} {\bibfnamefont {M.}~\bibnamefont {Razeghi}},\ }\bibfield  {title} {\bibinfo {title} {Background limited performance of long wavelength infrared focal plane arrays fabricated from m-structure inas–gasb superlattices},\ }\href {https://doi.org/10.1109/JQE.2008.2002667} {\bibfield  {journal} {\bibinfo  {journal} {IEEE J. Quantum Electron.}\ }\textbf {\bibinfo {volume} {45}},\ \bibinfo {pages} {157} (\bibinfo {year} {2009})}\BibitemShut {NoStop}%
\bibitem [{\citenamefont {Nguyen}\ \emph {et~al.}(2009)\citenamefont {Nguyen}, \citenamefont {Bogdanov}, \citenamefont {Pour},\ and\ \citenamefont {Razeghi}}]{2009NguyenAPL}%
  \BibitemOpen
  \bibfield  {author} {\bibinfo {author} {\bibfnamefont {B.-M.}\ \bibnamefont {Nguyen}}, \bibinfo {author} {\bibfnamefont {S.}~\bibnamefont {Bogdanov}}, \bibinfo {author} {\bibfnamefont {S.~A.}\ \bibnamefont {Pour}},\ and\ \bibinfo {author} {\bibfnamefont {M.}~\bibnamefont {Razeghi}},\ }\bibfield  {title} {\bibinfo {title} {Minority electron unipolar photodetectors based on type ii inas/gasb/alsb superlattices for very long wavelength infrared detection},\ }\href {https://doi.org/10.1063/1.3258489} {\bibfield  {journal} {\bibinfo  {journal} {Appl. Phys. Lett.}\ }\textbf {\bibinfo {volume} {95}},\ \bibinfo {pages} {183502} (\bibinfo {year} {2009})},\ \Eprint {https://arxiv.org/abs/https://doi.org/10.1063/1.3258489} {https://doi.org/10.1063/1.3258489} \BibitemShut {NoStop}%
\bibitem [{\citenamefont {Salihoglu}\ \emph {et~al.}(2012)\citenamefont {Salihoglu}, \citenamefont {Muti}, \citenamefont {Kutluer}, \citenamefont {Tansel}, \citenamefont {Turan}, \citenamefont {Ergun},\ and\ \citenamefont {Aydinli}}]{2012SalihogluAPL}%
  \BibitemOpen
  \bibfield  {author} {\bibinfo {author} {\bibfnamefont {O.}~\bibnamefont {Salihoglu}}, \bibinfo {author} {\bibfnamefont {A.}~\bibnamefont {Muti}}, \bibinfo {author} {\bibfnamefont {K.}~\bibnamefont {Kutluer}}, \bibinfo {author} {\bibfnamefont {T.}~\bibnamefont {Tansel}}, \bibinfo {author} {\bibfnamefont {R.}~\bibnamefont {Turan}}, \bibinfo {author} {\bibfnamefont {Y.}~\bibnamefont {Ergun}},\ and\ \bibinfo {author} {\bibfnamefont {A.}~\bibnamefont {Aydinli}},\ }\bibfield  {title} {\bibinfo {title} {“n” structure for type-ii superlattice photodetectors},\ }\href {https://doi.org/10.1063/1.4745841} {\bibfield  {journal} {\bibinfo  {journal} {Appl. Phys. Lett.}\ }\textbf {\bibinfo {volume} {101}},\ \bibinfo {pages} {073505} (\bibinfo {year} {2012})},\ \Eprint {https://arxiv.org/abs/https://doi.org/10.1063/1.4745841} {https://doi.org/10.1063/1.4745841} \BibitemShut {NoStop}%
\bibitem [{\citenamefont {Salihoglu}\ \emph {et~al.}(2014)\citenamefont {Salihoglu}, \citenamefont {Muti}, \citenamefont {Turan}, \citenamefont {Ergun},\ and\ \citenamefont {Aydinli}}]{2014Salihoglu}%
  \BibitemOpen
  \bibfield  {author} {\bibinfo {author} {\bibfnamefont {O.}~\bibnamefont {Salihoglu}}, \bibinfo {author} {\bibfnamefont {A.}~\bibnamefont {Muti}}, \bibinfo {author} {\bibfnamefont {R.}~\bibnamefont {Turan}}, \bibinfo {author} {\bibfnamefont {Y.}~\bibnamefont {Ergun}},\ and\ \bibinfo {author} {\bibfnamefont {A.}~\bibnamefont {Aydinli}},\ }\bibfield  {title} {\bibinfo {title} {{Low dark current N structure superlattice MWIR photodetectors}},\ }in\ \href {https://doi.org/10.1117/12.2050316} {\emph {\bibinfo {booktitle} {Infrared Technology and Applications XL}}},\ Vol.\ \bibinfo {volume} {9070},\ \bibinfo {editor} {edited by\ \bibinfo {editor} {\bibfnamefont {B.~F.}\ \bibnamefont {Andresen}}, \bibinfo {editor} {\bibfnamefont {G.~F.}\ \bibnamefont {Fulop}}, \bibinfo {editor} {\bibfnamefont {C.~M.}\ \bibnamefont {Hanson}},\ and\ \bibinfo {editor} {\bibfnamefont {P.~R.}\ \bibnamefont {Norton}}},\ \bibinfo {organization} {International Society for Optics and Photonics}\ (\bibinfo  {publisher} {SPIE},\ \bibinfo
  {year} {2014})\ p.\ \bibinfo {pages} {907012}\BibitemShut {NoStop}%
\bibitem [{\citenamefont {Hostut}\ \emph {et~al.}(2015)\citenamefont {Hostut}, \citenamefont {Alyoruk}, \citenamefont {Tansel}, \citenamefont {Kilic}, \citenamefont {Turan}, \citenamefont {Aydinli},\ and\ \citenamefont {Ergun}}]{2015HostutSLMS}%
  \BibitemOpen
  \bibfield  {author} {\bibinfo {author} {\bibfnamefont {M.}~\bibnamefont {Hostut}}, \bibinfo {author} {\bibfnamefont {M.}~\bibnamefont {Alyoruk}}, \bibinfo {author} {\bibfnamefont {T.}~\bibnamefont {Tansel}}, \bibinfo {author} {\bibfnamefont {A.}~\bibnamefont {Kilic}}, \bibinfo {author} {\bibfnamefont {R.}~\bibnamefont {Turan}}, \bibinfo {author} {\bibfnamefont {A.}~\bibnamefont {Aydinli}},\ and\ \bibinfo {author} {\bibfnamefont {Y.}~\bibnamefont {Ergun}},\ }\bibfield  {title} {\bibinfo {title} {N-structure based on inas/alsb/gasb superlattice photodetectors},\ }\href {https://doi.org/https://doi.org/10.1016/j.spmi.2014.12.022} {\bibfield  {journal} {\bibinfo  {journal} {Superlatt. Microstruct.}\ }\textbf {\bibinfo {volume} {79}},\ \bibinfo {pages} {116} (\bibinfo {year} {2015})}\BibitemShut {NoStop}%
\bibitem [{\citenamefont {Hinkey}\ and\ \citenamefont {Yang}(2013)}]{2013HinkeyJAP}%
  \BibitemOpen
  \bibfield  {author} {\bibinfo {author} {\bibfnamefont {R.~T.}\ \bibnamefont {Hinkey}}\ and\ \bibinfo {author} {\bibfnamefont {R.~Q.}\ \bibnamefont {Yang}},\ }\bibfield  {title} {\bibinfo {title} {Theory of multiple-stage interband photovoltaic devices and ultimate performance limit comparison of multiple-stage and single-stage interband infrared detectors},\ }\href {https://doi.org/10.1063/1.4820394} {\bibfield  {journal} {\bibinfo  {journal} {Journal of Applied Physics}\ }\textbf {\bibinfo {volume} {114}},\ \bibinfo {pages} {104506} (\bibinfo {year} {2013})},\ \Eprint {https://arxiv.org/abs/https://doi.org/10.1063/1.4820394} {https://doi.org/10.1063/1.4820394} \BibitemShut {NoStop}%
\bibitem [{\citenamefont {Pusz}\ \emph {et~al.}(2013)\citenamefont {Pusz}, \citenamefont {Kowalewski}, \citenamefont {Gawron}, \citenamefont {Plis}, \citenamefont {Krishna},\ and\ \citenamefont {Rogalski}}]{2013Pusz}%
  \BibitemOpen
  \bibfield  {author} {\bibinfo {author} {\bibfnamefont {W.}~\bibnamefont {Pusz}}, \bibinfo {author} {\bibfnamefont {A.}~\bibnamefont {Kowalewski}}, \bibinfo {author} {\bibfnamefont {W.}~\bibnamefont {Gawron}}, \bibinfo {author} {\bibfnamefont {E.}~\bibnamefont {Plis}}, \bibinfo {author} {\bibfnamefont {S.}~\bibnamefont {Krishna}},\ and\ \bibinfo {author} {\bibfnamefont {A.}~\bibnamefont {Rogalski}},\ }\bibfield  {title} {\bibinfo {title} {{MWIR type-II InAs/GaSb superlattice interband cascade photodetectors}},\ }in\ \href {https://doi.org/10.1117/12.2035499} {\emph {\bibinfo {booktitle} {Infrared Sensors, Devices, and Applications III}}},\ Vol.\ \bibinfo {volume} {8868},\ \bibinfo {editor} {edited by\ \bibinfo {editor} {\bibfnamefont {P.~D.}\ \bibnamefont {LeVan}}, \bibinfo {editor} {\bibfnamefont {A.~K.}\ \bibnamefont {Sood}}, \bibinfo {editor} {\bibfnamefont {P.~S.}\ \bibnamefont {Wijewarnasuriya}},\ and\ \bibinfo {editor} {\bibfnamefont {A.~I.}\ \bibnamefont {D'Souza}}},\ \bibinfo {organization}
  {International Society for Optics and Photonics}\ (\bibinfo  {publisher} {SPIE},\ \bibinfo {year} {2013})\ p.\ \bibinfo {pages} {88680M}\BibitemShut {NoStop}%
\bibitem [{\citenamefont {Gawron}\ \emph {et~al.}(2022)\citenamefont {Gawron}, \citenamefont {Kubiszyn}, \citenamefont {Michalczewski}, \citenamefont {Piotrowski},\ and\ \citenamefont {Martyniuk}}]{2022GawronEDL}%
  \BibitemOpen
  \bibfield  {author} {\bibinfo {author} {\bibfnamefont {W.}~\bibnamefont {Gawron}}, \bibinfo {author} {\bibfnamefont {{\L}.}~\bibnamefont {Kubiszyn}}, \bibinfo {author} {\bibfnamefont {K.}~\bibnamefont {Michalczewski}}, \bibinfo {author} {\bibfnamefont {J.}~\bibnamefont {Piotrowski}},\ and\ \bibinfo {author} {\bibfnamefont {P.}~\bibnamefont {Martyniuk}},\ }\bibfield  {title} {\bibinfo {title} {Demonstration of the longwave type-ii superlattice inas/inassb cascade photodetector for high operating temperature},\ }\href {https://doi.org/10.1109/LED.2022.3188909} {\bibfield  {journal} {\bibinfo  {journal} {IEEE Electron Device Letters}\ }\textbf {\bibinfo {volume} {43}},\ \bibinfo {pages} {1487} (\bibinfo {year} {2022})}\BibitemShut {NoStop}%
\bibitem [{\citenamefont {Martyniuk}\ \emph {et~al.}(2022)\citenamefont {Martyniuk}, \citenamefont {Rogalski},\ and\ \citenamefont {Krishna}}]{2022MartyniukPRAPP}%
  \BibitemOpen
  \bibfield  {author} {\bibinfo {author} {\bibfnamefont {P.}~\bibnamefont {Martyniuk}}, \bibinfo {author} {\bibfnamefont {A.}~\bibnamefont {Rogalski}},\ and\ \bibinfo {author} {\bibfnamefont {S.}~\bibnamefont {Krishna}},\ }\bibfield  {title} {\bibinfo {title} {Interband quantum cascade infrared photodetectors: Current status and future trends},\ }\href {https://doi.org/10.1103/PhysRevApplied.17.027001} {\bibfield  {journal} {\bibinfo  {journal} {Phys. Rev. Appl.}\ }\textbf {\bibinfo {volume} {17}},\ \bibinfo {pages} {027001} (\bibinfo {year} {2022})}\BibitemShut {NoStop}%
\bibitem [{\citenamefont {Höglund}\ \emph {et~al.}(2013)\citenamefont {Höglund}, \citenamefont {Ting}, \citenamefont {Khoshakhlagh}, \citenamefont {Soibel}, \citenamefont {Hill}, \citenamefont {Fisher}, \citenamefont {Keo},\ and\ \citenamefont {Gunapala}}]{2013HoglundAPL}%
  \BibitemOpen
  \bibfield  {author} {\bibinfo {author} {\bibfnamefont {L.}~\bibnamefont {Höglund}}, \bibinfo {author} {\bibfnamefont {D.~Z.}\ \bibnamefont {Ting}}, \bibinfo {author} {\bibfnamefont {A.}~\bibnamefont {Khoshakhlagh}}, \bibinfo {author} {\bibfnamefont {A.}~\bibnamefont {Soibel}}, \bibinfo {author} {\bibfnamefont {C.~J.}\ \bibnamefont {Hill}}, \bibinfo {author} {\bibfnamefont {A.}~\bibnamefont {Fisher}}, \bibinfo {author} {\bibfnamefont {S.}~\bibnamefont {Keo}},\ and\ \bibinfo {author} {\bibfnamefont {S.~D.}\ \bibnamefont {Gunapala}},\ }\bibfield  {title} {\bibinfo {title} {Influence of radiative and non-radiative recombination on the minority carrier lifetime in midwave infrared inas/inassb superlattices},\ }\href {https://doi.org/10.1063/1.4835055} {\bibfield  {journal} {\bibinfo  {journal} {Appl. Phys. Lett.}\ }\textbf {\bibinfo {volume} {103}},\ \bibinfo {pages} {221908} (\bibinfo {year} {2013})},\ \Eprint {https://arxiv.org/abs/https://doi.org/10.1063/1.4835055} {https://doi.org/10.1063/1.4835055}
  \BibitemShut {NoStop}%
\bibitem [{\citenamefont {Steenbergen}\ \emph {et~al.}(2011)\citenamefont {Steenbergen}, \citenamefont {Connelly}, \citenamefont {Metcalfe}, \citenamefont {Shen}, \citenamefont {Wraback}, \citenamefont {Lubyshev}, \citenamefont {Qiu}, \citenamefont {Fastenau}, \citenamefont {Liu}, \citenamefont {Elhamri}, \citenamefont {Cellek},\ and\ \citenamefont {Zhang}}]{2011SteenbergenAPL}%
  \BibitemOpen
  \bibfield  {author} {\bibinfo {author} {\bibfnamefont {E.~H.}\ \bibnamefont {Steenbergen}}, \bibinfo {author} {\bibfnamefont {B.~C.}\ \bibnamefont {Connelly}}, \bibinfo {author} {\bibfnamefont {G.~D.}\ \bibnamefont {Metcalfe}}, \bibinfo {author} {\bibfnamefont {H.}~\bibnamefont {Shen}}, \bibinfo {author} {\bibfnamefont {M.}~\bibnamefont {Wraback}}, \bibinfo {author} {\bibfnamefont {D.}~\bibnamefont {Lubyshev}}, \bibinfo {author} {\bibfnamefont {Y.}~\bibnamefont {Qiu}}, \bibinfo {author} {\bibfnamefont {J.~M.}\ \bibnamefont {Fastenau}}, \bibinfo {author} {\bibfnamefont {A.~W.~K.}\ \bibnamefont {Liu}}, \bibinfo {author} {\bibfnamefont {S.}~\bibnamefont {Elhamri}}, \bibinfo {author} {\bibfnamefont {O.~O.}\ \bibnamefont {Cellek}},\ and\ \bibinfo {author} {\bibfnamefont {Y.-H.}\ \bibnamefont {Zhang}},\ }\bibfield  {title} {\bibinfo {title} {Significantly improved minority carrier lifetime observed in a long-wavelength infrared iii-v type-ii superlattice comprised of inas/inassb},\ }\href
  {https://doi.org/10.1063/1.3671398} {\bibfield  {journal} {\bibinfo  {journal} {Appl. Phys. Lett.}\ }\textbf {\bibinfo {volume} {99}},\ \bibinfo {pages} {251110} (\bibinfo {year} {2011})},\ \Eprint {https://arxiv.org/abs/https://doi.org/10.1063/1.3671398} {https://doi.org/10.1063/1.3671398} \BibitemShut {NoStop}%
\bibitem [{\citenamefont {Ting}\ \emph {et~al.}(2020)\citenamefont {Ting}, \citenamefont {Rafol}, \citenamefont {Khoshakhlagh}, \citenamefont {Soibel}, \citenamefont {Keo}, \citenamefont {Fisher}, \citenamefont {Pepper}, \citenamefont {Hill},\ and\ \citenamefont {Gunapala}}]{2020TingMMA}%
  \BibitemOpen
  \bibfield  {author} {\bibinfo {author} {\bibfnamefont {D.~Z.}\ \bibnamefont {Ting}}, \bibinfo {author} {\bibfnamefont {S.~B.}\ \bibnamefont {Rafol}}, \bibinfo {author} {\bibfnamefont {A.}~\bibnamefont {Khoshakhlagh}}, \bibinfo {author} {\bibfnamefont {A.}~\bibnamefont {Soibel}}, \bibinfo {author} {\bibfnamefont {S.~A.}\ \bibnamefont {Keo}}, \bibinfo {author} {\bibfnamefont {A.~M.}\ \bibnamefont {Fisher}}, \bibinfo {author} {\bibfnamefont {B.~J.}\ \bibnamefont {Pepper}}, \bibinfo {author} {\bibfnamefont {C.~J.}\ \bibnamefont {Hill}},\ and\ \bibinfo {author} {\bibfnamefont {S.~D.}\ \bibnamefont {Gunapala}},\ }\bibfield  {title} {\bibinfo {title} {Inas/inassb type-ii strained-layer superlattice infrared photodetectors},\ }\bibfield  {journal} {\bibinfo  {journal} {Micromachines}\ }\textbf {\bibinfo {volume} {11}},\ \href {https://doi.org/10.3390/mi11110958} {10.3390/mi11110958} (\bibinfo {year} {2020})\BibitemShut {NoStop}%
\bibitem [{\citenamefont {Olson}\ \emph {et~al.}(2017)\citenamefont {Olson}, \citenamefont {Klem}, \citenamefont {Kadlec}, \citenamefont {Kim}, \citenamefont {Goldflam}, \citenamefont {Hawkins}, \citenamefont {Tauke-Pedretti}, \citenamefont {Coon}, \citenamefont {Fortune}, \citenamefont {Shaner},\ and\ \citenamefont {Flatt\'e}}]{2017OlsonPRA}%
  \BibitemOpen
  \bibfield  {author} {\bibinfo {author} {\bibfnamefont {B.~V.}\ \bibnamefont {Olson}}, \bibinfo {author} {\bibfnamefont {J.~F.}\ \bibnamefont {Klem}}, \bibinfo {author} {\bibfnamefont {E.~A.}\ \bibnamefont {Kadlec}}, \bibinfo {author} {\bibfnamefont {J.~K.}\ \bibnamefont {Kim}}, \bibinfo {author} {\bibfnamefont {M.~D.}\ \bibnamefont {Goldflam}}, \bibinfo {author} {\bibfnamefont {S.~D.}\ \bibnamefont {Hawkins}}, \bibinfo {author} {\bibfnamefont {A.}~\bibnamefont {Tauke-Pedretti}}, \bibinfo {author} {\bibfnamefont {W.~T.}\ \bibnamefont {Coon}}, \bibinfo {author} {\bibfnamefont {T.~R.}\ \bibnamefont {Fortune}}, \bibinfo {author} {\bibfnamefont {E.~A.}\ \bibnamefont {Shaner}},\ and\ \bibinfo {author} {\bibfnamefont {M.~E.}\ \bibnamefont {Flatt\'e}},\ }\bibfield  {title} {\bibinfo {title} {Vertical hole transport and carrier localization in $\mathrm{InAs}/{\mathrm{inas}}_{1\ensuremath{-}x}{\mathrm{sb}}_{x}$ type-ii superlattice heterojunction bipolar transistors},\ }\href
  {https://doi.org/10.1103/PhysRevApplied.7.024016} {\bibfield  {journal} {\bibinfo  {journal} {Phys. Rev. Appl.}\ }\textbf {\bibinfo {volume} {7}},\ \bibinfo {pages} {024016} (\bibinfo {year} {2017})}\BibitemShut {NoStop}%
\bibitem [{\citenamefont {Casias}\ \emph {et~al.}(2020)\citenamefont {Casias}, \citenamefont {Morath}, \citenamefont {Steenbergen}, \citenamefont {Umana-Membreno}, \citenamefont {Webster}, \citenamefont {Logan}, \citenamefont {Kim}, \citenamefont {Balakrishnan}, \citenamefont {Faraone},\ and\ \citenamefont {Krishna}}]{2020CasiasAPL}%
  \BibitemOpen
  \bibfield  {author} {\bibinfo {author} {\bibfnamefont {L.~K.}\ \bibnamefont {Casias}}, \bibinfo {author} {\bibfnamefont {C.~P.}\ \bibnamefont {Morath}}, \bibinfo {author} {\bibfnamefont {E.~H.}\ \bibnamefont {Steenbergen}}, \bibinfo {author} {\bibfnamefont {G.~A.}\ \bibnamefont {Umana-Membreno}}, \bibinfo {author} {\bibfnamefont {P.~T.}\ \bibnamefont {Webster}}, \bibinfo {author} {\bibfnamefont {J.~V.}\ \bibnamefont {Logan}}, \bibinfo {author} {\bibfnamefont {J.~K.}\ \bibnamefont {Kim}}, \bibinfo {author} {\bibfnamefont {G.}~\bibnamefont {Balakrishnan}}, \bibinfo {author} {\bibfnamefont {L.}~\bibnamefont {Faraone}},\ and\ \bibinfo {author} {\bibfnamefont {S.}~\bibnamefont {Krishna}},\ }\bibfield  {title} {\bibinfo {title} {Vertical carrier transport in strain-balanced inas/inassb type-ii superlattice material},\ }\href {https://doi.org/10.1063/1.5144079} {\bibfield  {journal} {\bibinfo  {journal} {Appl. Phys. Lett.}\ }\textbf {\bibinfo {volume} {116}},\ \bibinfo {pages} {182109} (\bibinfo {year} {2020})},\
  \Eprint {https://arxiv.org/abs/https://doi.org/10.1063/1.5144079} {https://doi.org/10.1063/1.5144079} \BibitemShut {NoStop}%
\bibitem [{\citenamefont {Bellotti}\ \emph {et~al.}(2021)\citenamefont {Bellotti}, \citenamefont {Bertazzi}, \citenamefont {Tibaldi}, \citenamefont {Schuster}, \citenamefont {Bajaj},\ and\ \citenamefont {Reed}}]{2021BellottiPRAPP}%
  \BibitemOpen
  \bibfield  {author} {\bibinfo {author} {\bibfnamefont {E.}~\bibnamefont {Bellotti}}, \bibinfo {author} {\bibfnamefont {F.}~\bibnamefont {Bertazzi}}, \bibinfo {author} {\bibfnamefont {A.}~\bibnamefont {Tibaldi}}, \bibinfo {author} {\bibfnamefont {J.}~\bibnamefont {Schuster}}, \bibinfo {author} {\bibfnamefont {J.}~\bibnamefont {Bajaj}},\ and\ \bibinfo {author} {\bibfnamefont {M.}~\bibnamefont {Reed}},\ }\bibfield  {title} {\bibinfo {title} {Disorder-induced degradation of vertical carrier transport in strain-balanced antimony-based superlattices},\ }\href {https://doi.org/10.1103/PhysRevApplied.16.054028} {\bibfield  {journal} {\bibinfo  {journal} {Phys. Rev. Appl.}\ }\textbf {\bibinfo {volume} {16}},\ \bibinfo {pages} {054028} (\bibinfo {year} {2021})}\BibitemShut {NoStop}%
\bibitem [{\citenamefont {Tsu}\ and\ \citenamefont {Esaki}(1991)}]{1991TsuPRB}%
  \BibitemOpen
  \bibfield  {author} {\bibinfo {author} {\bibfnamefont {R.}~\bibnamefont {Tsu}}\ and\ \bibinfo {author} {\bibfnamefont {L.}~\bibnamefont {Esaki}},\ }\bibfield  {title} {\bibinfo {title} {Stark quantization in superlattices},\ }\href {https://doi.org/10.1103/PhysRevB.43.5204} {\bibfield  {journal} {\bibinfo  {journal} {Phys. Rev. B}\ }\textbf {\bibinfo {volume} {43}},\ \bibinfo {pages} {5204} (\bibinfo {year} {1991})}\BibitemShut {NoStop}%
\bibitem [{\citenamefont {Laikhtman}\ and\ \citenamefont {Miller}(1993)}]{1993LaikhtmanPRB}%
  \BibitemOpen
  \bibfield  {author} {\bibinfo {author} {\bibfnamefont {B.}~\bibnamefont {Laikhtman}}\ and\ \bibinfo {author} {\bibfnamefont {D.}~\bibnamefont {Miller}},\ }\bibfield  {title} {\bibinfo {title} {Theory of current-voltage instabilities in superlattices},\ }\href {https://doi.org/10.1103/PhysRevB.48.5395} {\bibfield  {journal} {\bibinfo  {journal} {Phys. Rev. B}\ }\textbf {\bibinfo {volume} {48}},\ \bibinfo {pages} {5395} (\bibinfo {year} {1993})}\BibitemShut {NoStop}%
\bibitem [{\citenamefont {Wacker}\ and\ \citenamefont {Jauho}(1998)}]{1998WackerPRL}%
  \BibitemOpen
  \bibfield  {author} {\bibinfo {author} {\bibfnamefont {A.}~\bibnamefont {Wacker}}\ and\ \bibinfo {author} {\bibfnamefont {A.-P.}\ \bibnamefont {Jauho}},\ }\bibfield  {title} {\bibinfo {title} {Quantum transport: The link between standard approaches in superlattices},\ }\href {https://doi.org/10.1103/PhysRevLett.80.369} {\bibfield  {journal} {\bibinfo  {journal} {Phys. Rev. Lett.}\ }\textbf {\bibinfo {volume} {80}},\ \bibinfo {pages} {369} (\bibinfo {year} {1998})}\BibitemShut {NoStop}%
\bibitem [{\citenamefont {Bertazzi}\ \emph {et~al.}(2020)\citenamefont {Bertazzi}, \citenamefont {Tibaldi}, \citenamefont {Goano}, \citenamefont {Montoya},\ and\ \citenamefont {Bellotti}}]{2020BertazziPRAPP}%
  \BibitemOpen
  \bibfield  {author} {\bibinfo {author} {\bibfnamefont {F.}~\bibnamefont {Bertazzi}}, \bibinfo {author} {\bibfnamefont {A.}~\bibnamefont {Tibaldi}}, \bibinfo {author} {\bibfnamefont {M.}~\bibnamefont {Goano}}, \bibinfo {author} {\bibfnamefont {J.~A.~G.}\ \bibnamefont {Montoya}},\ and\ \bibinfo {author} {\bibfnamefont {E.}~\bibnamefont {Bellotti}},\ }\bibfield  {title} {\bibinfo {title} {Nonequilibrium green's function modeling of type-ii superlattice detectors and its connection to semiclassical approaches},\ }\href {https://doi.org/10.1103/PhysRevApplied.14.014083} {\bibfield  {journal} {\bibinfo  {journal} {Phys. Rev. Appl.}\ }\textbf {\bibinfo {volume} {14}},\ \bibinfo {pages} {014083} (\bibinfo {year} {2020})}\BibitemShut {NoStop}%
\bibitem [{\citenamefont {Glennon}\ \emph {et~al.}(2023)\citenamefont {Glennon}, \citenamefont {Bertazzi}, \citenamefont {Tibaldi},\ and\ \citenamefont {Bellotti}}]{2023GlennonPRAPP}%
  \BibitemOpen
  \bibfield  {author} {\bibinfo {author} {\bibfnamefont {J.}~\bibnamefont {Glennon}}, \bibinfo {author} {\bibfnamefont {F.}~\bibnamefont {Bertazzi}}, \bibinfo {author} {\bibfnamefont {A.}~\bibnamefont {Tibaldi}},\ and\ \bibinfo {author} {\bibfnamefont {E.}~\bibnamefont {Bellotti}},\ }\bibfield  {title} {\bibinfo {title} {Extraction of mobility from quantum transport calculations of type-ii superlattices},\ }\href {https://doi.org/10.1103/PhysRevApplied.19.044045} {\bibfield  {journal} {\bibinfo  {journal} {Phys. Rev. Appl.}\ }\textbf {\bibinfo {volume} {19}},\ \bibinfo {pages} {044045} (\bibinfo {year} {2023})}\BibitemShut {NoStop}%
\bibitem [{\citenamefont {Smith}(2009)}]{Abaqus}%
  \BibitemOpen
  \bibfield  {author} {\bibinfo {author} {\bibfnamefont {M.}~\bibnamefont {Smith}},\ }\href@noop {} {{\selectlanguage {English}\emph {\bibinfo {title} {ABAQUS/Standard User's Manual, Version 6.9}}}}\ (\bibinfo  {publisher} {Dassault Syst{\`e}mes Simulia Corp},\ \bibinfo {address} {United States},\ \bibinfo {year} {2009})\BibitemShut {NoStop}%
\bibitem [{\citenamefont {Qiao}\ \emph {et~al.}(2012)\citenamefont {Qiao}, \citenamefont {Mou},\ and\ \citenamefont {Chuang}}]{2012QiaoOEX}%
  \BibitemOpen
  \bibfield  {author} {\bibinfo {author} {\bibfnamefont {P.-F.}\ \bibnamefont {Qiao}}, \bibinfo {author} {\bibfnamefont {S.}~\bibnamefont {Mou}},\ and\ \bibinfo {author} {\bibfnamefont {S.~L.}\ \bibnamefont {Chuang}},\ }\bibfield  {title} {\bibinfo {title} {Electronic band structures and optical properties of type-ii superlattice photodetectors with interfacial effect},\ }\href {https://doi.org/10.1364/OE.20.002319} {\bibfield  {journal} {\bibinfo  {journal} {Opt. Express}\ }\textbf {\bibinfo {volume} {20}},\ \bibinfo {pages} {2319} (\bibinfo {year} {2012})}\BibitemShut {NoStop}%
\bibitem [{\citenamefont {Webster}\ \emph {et~al.}(2015)\citenamefont {Webster}, \citenamefont {Riordan}, \citenamefont {Liu}, \citenamefont {Steenbergen}, \citenamefont {Synowicki}, \citenamefont {Zhang},\ and\ \citenamefont {Johnson}}]{2015WebsterJAP}%
  \BibitemOpen
  \bibfield  {author} {\bibinfo {author} {\bibfnamefont {P.~T.}\ \bibnamefont {Webster}}, \bibinfo {author} {\bibfnamefont {N.~A.}\ \bibnamefont {Riordan}}, \bibinfo {author} {\bibfnamefont {S.}~\bibnamefont {Liu}}, \bibinfo {author} {\bibfnamefont {E.~H.}\ \bibnamefont {Steenbergen}}, \bibinfo {author} {\bibfnamefont {R.~A.}\ \bibnamefont {Synowicki}}, \bibinfo {author} {\bibfnamefont {Y.-H.}\ \bibnamefont {Zhang}},\ and\ \bibinfo {author} {\bibfnamefont {S.~R.}\ \bibnamefont {Johnson}},\ }\bibfield  {title} {\bibinfo {title} {{Measurement of InAsSb bandgap energy and InAs/InAsSb band edge positions using spectroscopic ellipsometry and photoluminescence spectroscopy}},\ }\bibfield  {journal} {\bibinfo  {journal} {J. Appl. Phys.}\ }\textbf {\bibinfo {volume} {118}},\ \href {https://doi.org/10.1063/1.4939293} {10.1063/1.4939293} (\bibinfo {year} {2015}),\ \bibinfo {note} {245706}\BibitemShut {NoStop}%
\bibitem [{\citenamefont {Enders}\ \emph {et~al.}(1995)\citenamefont {Enders}, \citenamefont {B\"arwolff}, \citenamefont {Woerner},\ and\ \citenamefont {Suisky}}]{1995EndersPRB}%
  \BibitemOpen
  \bibfield  {author} {\bibinfo {author} {\bibfnamefont {P.}~\bibnamefont {Enders}}, \bibinfo {author} {\bibfnamefont {A.}~\bibnamefont {B\"arwolff}}, \bibinfo {author} {\bibfnamefont {M.}~\bibnamefont {Woerner}},\ and\ \bibinfo {author} {\bibfnamefont {D.}~\bibnamefont {Suisky}},\ }\bibfield  {title} {\bibinfo {title} {k\ensuremath{\cdot}p theory of energy bands, wave functions, and optical selection rules in strained tetrahedral semiconductors},\ }\href {https://doi.org/10.1103/PhysRevB.51.16695} {\bibfield  {journal} {\bibinfo  {journal} {Phys. Rev. B}\ }\textbf {\bibinfo {volume} {51}},\ \bibinfo {pages} {16695} (\bibinfo {year} {1995})}\BibitemShut {NoStop}%
\bibitem [{\citenamefont {Muraki}\ \emph {et~al.}(1992)\citenamefont {Muraki}, \citenamefont {Fukatsu}, \citenamefont {Shiraki},\ and\ \citenamefont {Ito}}]{1992MurakiAPL}%
  \BibitemOpen
  \bibfield  {author} {\bibinfo {author} {\bibfnamefont {K.}~\bibnamefont {Muraki}}, \bibinfo {author} {\bibfnamefont {S.}~\bibnamefont {Fukatsu}}, \bibinfo {author} {\bibfnamefont {Y.}~\bibnamefont {Shiraki}},\ and\ \bibinfo {author} {\bibfnamefont {R.}~\bibnamefont {Ito}},\ }\bibfield  {title} {\bibinfo {title} {Surface segregation of in atoms during molecular beam epitaxy and its influence on the energy levels in ingosbournaas/gaas quantum wells},\ }\href {https://doi.org/10.1063/1.107835} {\bibfield  {journal} {\bibinfo  {journal} {Appl. Phys. Lett.}\ }\textbf {\bibinfo {volume} {61}},\ \bibinfo {pages} {557} (\bibinfo {year} {1992})},\ \Eprint {https://arxiv.org/abs/https://doi.org/10.1063/1.107835} {https://doi.org/10.1063/1.107835} \BibitemShut {NoStop}%
\bibitem [{\citenamefont {Kim}\ \emph {et~al.}(2018)\citenamefont {Kim}, \citenamefont {Meng}, \citenamefont {Klem}, \citenamefont {Hawkins}, \citenamefont {Kim},\ and\ \citenamefont {Zuo}}]{2018KimJAP}%
  \BibitemOpen
  \bibfield  {author} {\bibinfo {author} {\bibfnamefont {H.}~\bibnamefont {Kim}}, \bibinfo {author} {\bibfnamefont {Y.}~\bibnamefont {Meng}}, \bibinfo {author} {\bibfnamefont {J.~F.}\ \bibnamefont {Klem}}, \bibinfo {author} {\bibfnamefont {S.~D.}\ \bibnamefont {Hawkins}}, \bibinfo {author} {\bibfnamefont {J.~K.}\ \bibnamefont {Kim}},\ and\ \bibinfo {author} {\bibfnamefont {J.-M.}\ \bibnamefont {Zuo}},\ }\bibfield  {title} {\bibinfo {title} {Sb-induced strain fluctuations in a strained layer superlattice of inas/inassb},\ }\href {https://doi.org/10.1063/1.4993673} {\bibfield  {journal} {\bibinfo  {journal} {J. Appl. Phys.}\ }\textbf {\bibinfo {volume} {123}},\ \bibinfo {pages} {161521} (\bibinfo {year} {2018})},\ \Eprint {https://arxiv.org/abs/https://doi.org/10.1063/1.4993673} {https://doi.org/10.1063/1.4993673} \BibitemShut {NoStop}%
\bibitem [{\citenamefont {Luisier}\ \emph {et~al.}(2006)\citenamefont {Luisier}, \citenamefont {Schenk}, \citenamefont {Fichtner},\ and\ \citenamefont {Klimeck}}]{2006LuisierPRB}%
  \BibitemOpen
  \bibfield  {author} {\bibinfo {author} {\bibfnamefont {M.}~\bibnamefont {Luisier}}, \bibinfo {author} {\bibfnamefont {A.}~\bibnamefont {Schenk}}, \bibinfo {author} {\bibfnamefont {W.}~\bibnamefont {Fichtner}},\ and\ \bibinfo {author} {\bibfnamefont {G.}~\bibnamefont {Klimeck}},\ }\bibfield  {title} {\bibinfo {title} {Atomistic simulation of nanowires in the $s{p}^{3}{d}^{5}{s}^{*}$ tight-binding formalism: From boundary conditions to strain calculations},\ }\href {https://doi.org/10.1103/PhysRevB.74.205323} {\bibfield  {journal} {\bibinfo  {journal} {Phys. Rev. B}\ }\textbf {\bibinfo {volume} {74}},\ \bibinfo {pages} {205323} (\bibinfo {year} {2006})}\BibitemShut {NoStop}%
\bibitem [{\citenamefont {Aeberhard}(2008)}]{2008Aeberhard_PhD}%
  \BibitemOpen
  \bibfield  {author} {\bibinfo {author} {\bibfnamefont {U.}~\bibnamefont {Aeberhard}},\ }\emph {\bibinfo {title} {A Microscopic Theory of Quantum Well Photovoltaics}},\ \href {https://doi.org/10.3929/ethz-a-005761970} {Ph.D. thesis},\ \bibinfo  {school} {Eidgen{\"o}ssische Technische Hochschule Z{\"u}rich} (\bibinfo {year} {2008})\BibitemShut {NoStop}%
\bibitem [{\citenamefont {Steiger}(2009)}]{2009Steiger_PhD}%
  \BibitemOpen
  \bibfield  {author} {\bibinfo {author} {\bibfnamefont {S.}~\bibnamefont {Steiger}},\ }\emph {\bibinfo {title} {Modelling Nano-{LED}s}},\ \href {https://doi.org/10.3929/ethz-a-005834599} {Ph.D. thesis},\ \bibinfo  {school} {Eidgen{\"o}ssische Technische Hochschule Z{\"u}rich} (\bibinfo {year} {2009})\BibitemShut {NoStop}%
\bibitem [{\citenamefont {Pedregosa}\ \emph {et~al.}(2011)\citenamefont {Pedregosa}, \citenamefont {Varoquaux}, \citenamefont {Gramfort}, \citenamefont {Michel}, \citenamefont {Thirion}, \citenamefont {Grisel}, \citenamefont {Blondel}, \citenamefont {Prettenhofer}, \citenamefont {Weiss}, \citenamefont {Dubourg}, \citenamefont {Vanderplas}, \citenamefont {Passos}, \citenamefont {Cournapeau}, \citenamefont {Brucher}, \citenamefont {Perrot},\ and\ \citenamefont {Duchesnay}}]{pedregosa2011scikitlearn}%
  \BibitemOpen
  \bibfield  {author} {\bibinfo {author} {\bibfnamefont {F.}~\bibnamefont {Pedregosa}}, \bibinfo {author} {\bibfnamefont {G.}~\bibnamefont {Varoquaux}}, \bibinfo {author} {\bibfnamefont {A.}~\bibnamefont {Gramfort}}, \bibinfo {author} {\bibfnamefont {V.}~\bibnamefont {Michel}}, \bibinfo {author} {\bibfnamefont {B.}~\bibnamefont {Thirion}}, \bibinfo {author} {\bibfnamefont {O.}~\bibnamefont {Grisel}}, \bibinfo {author} {\bibfnamefont {M.}~\bibnamefont {Blondel}}, \bibinfo {author} {\bibfnamefont {P.}~\bibnamefont {Prettenhofer}}, \bibinfo {author} {\bibfnamefont {R.}~\bibnamefont {Weiss}}, \bibinfo {author} {\bibfnamefont {V.}~\bibnamefont {Dubourg}}, \bibinfo {author} {\bibfnamefont {J.}~\bibnamefont {Vanderplas}}, \bibinfo {author} {\bibfnamefont {A.}~\bibnamefont {Passos}}, \bibinfo {author} {\bibfnamefont {D.}~\bibnamefont {Cournapeau}}, \bibinfo {author} {\bibfnamefont {M.}~\bibnamefont {Brucher}}, \bibinfo {author} {\bibfnamefont {M.}~\bibnamefont {Perrot}},\ and\ \bibinfo {author} {\bibfnamefont
  {E.}~\bibnamefont {Duchesnay}},\ }\bibfield  {title} {\bibinfo {title} {Scikit-learn: Machine learning in {P}ython},\ }\href {http://www.jmlr.org/papers/volume12/pedregosa11a/pedregosa11a.pdf} {\bibfield  {journal} {\bibinfo  {journal} {Journal of Machine Learning Research}\ }\textbf {\bibinfo {volume} {12}},\ \bibinfo {pages} {2825} (\bibinfo {year} {2011})}\BibitemShut {NoStop}%
\bibitem [{\citenamefont {Reine}\ \emph {et~al.}(1981)\citenamefont {Reine}, \citenamefont {Sood},\ and\ \citenamefont {Tredwell}}]{1981ReineSCSM}%
  \BibitemOpen
  \bibfield  {author} {\bibinfo {author} {\bibfnamefont {M.}~\bibnamefont {Reine}}, \bibinfo {author} {\bibfnamefont {A.}~\bibnamefont {Sood}},\ and\ \bibinfo {author} {\bibfnamefont {T.}~\bibnamefont {Tredwell}},\ }\bibfield  {title} {\bibinfo {title} {Chapter 6 photovoltaic infrared detectors},\ }in\ \href {https://doi.org/https://doi.org/10.1016/S0080-8784(08)62766-0} {\emph {\bibinfo {booktitle} {Mercury Cadmium Telluride}}},\ \bibinfo {series} {Semiconductors and Semimetals}, Vol.~\bibinfo {volume} {18},\ \bibinfo {editor} {edited by\ \bibinfo {editor} {\bibfnamefont {R.}~\bibnamefont {Willardson}}\ and\ \bibinfo {editor} {\bibfnamefont {A.~C.}\ \bibnamefont {Beer}}}\ (\bibinfo  {publisher} {Elsevier},\ \bibinfo {year} {1981})\ pp.\ \bibinfo {pages} {201--311}\BibitemShut {NoStop}%
\bibitem [{MAT(2020)}]{MATLAB:R2020b_u5}%
  \BibitemOpen
  \href@noop {} {\emph {\bibinfo {title} {{MATLAB version 9.9.0.1592791 (R2020b) Update 5}}}},\ \bibinfo {organization} {The Mathworks, Inc.},\ \bibinfo {address} {Natick, Massachusetts} (\bibinfo {year} {2020})\BibitemShut {NoStop}%
\bibitem [{\citenamefont {Aalok}(2021)}]{atharva2021}%
  \BibitemOpen
  \bibfield  {author} {\bibinfo {author} {\bibfnamefont {A.}~\bibnamefont {Aalok}},\ }\href@noop {} {\bibinfo {title} {Professional plots}},\ \bibinfo {howpublished} {\url{https://www.mathworks.com/matlabcentral/fileexchange/100766-~professional_plots}} (\bibinfo {year} {2021}),\ \bibinfo {note} {[Online; accessed October 31, 2022]}\BibitemShut {NoStop}%
\bibitem [{\citenamefont {Ting}\ \emph {et~al.}(2016)\citenamefont {Ting}, \citenamefont {Soibel},\ and\ \citenamefont {Gunapala}}]{2016TingAPL}%
  \BibitemOpen
  \bibfield  {author} {\bibinfo {author} {\bibfnamefont {D.~Z.}\ \bibnamefont {Ting}}, \bibinfo {author} {\bibfnamefont {A.}~\bibnamefont {Soibel}},\ and\ \bibinfo {author} {\bibfnamefont {S.~D.}\ \bibnamefont {Gunapala}},\ }\bibfield  {title} {\bibinfo {title} {Hole effective masses and subband splitting in type-ii superlattice infrared detectors},\ }\href {https://doi.org/10.1063/1.4948387} {\bibfield  {journal} {\bibinfo  {journal} {Appl. Phys. Lett.}\ }\textbf {\bibinfo {volume} {108}},\ \bibinfo {pages} {183504} (\bibinfo {year} {2016})},\ \Eprint {https://arxiv.org/abs/https://doi.org/10.1063/1.4948387} {https://doi.org/10.1063/1.4948387} \BibitemShut {NoStop}%
\bibitem [{\citenamefont {Liu}\ \emph {et~al.}(2022)\citenamefont {Liu}, \citenamefont {Donetski}, \citenamefont {Kucharczyk}, \citenamefont {Zhao}, \citenamefont {Kipshidze}, \citenamefont {Belenky},\ and\ \citenamefont {Svensson}}]{2022LiuAPL}%
  \BibitemOpen
  \bibfield  {author} {\bibinfo {author} {\bibfnamefont {J.}~\bibnamefont {Liu}}, \bibinfo {author} {\bibfnamefont {D.}~\bibnamefont {Donetski}}, \bibinfo {author} {\bibfnamefont {K.}~\bibnamefont {Kucharczyk}}, \bibinfo {author} {\bibfnamefont {J.}~\bibnamefont {Zhao}}, \bibinfo {author} {\bibfnamefont {G.}~\bibnamefont {Kipshidze}}, \bibinfo {author} {\bibfnamefont {G.}~\bibnamefont {Belenky}},\ and\ \bibinfo {author} {\bibfnamefont {S.~P.}\ \bibnamefont {Svensson}},\ }\bibfield  {title} {\bibinfo {title} {{Short-period InAsSb-based strained layer superlattices for high quantum efficiency long-wave infrared detectors}},\ }\bibfield  {journal} {\bibinfo  {journal} {Appl. Phys. Lett.}\ }\textbf {\bibinfo {volume} {120}},\ \href {https://doi.org/10.1063/5.0083862} {10.1063/5.0083862} (\bibinfo {year} {2022}),\ \bibinfo {note} {141101}\BibitemShut {NoStop}%
\bibitem [{\citenamefont {Donetsky}\ \emph {et~al.}(2010)\citenamefont {Donetsky}, \citenamefont {Belenky}, \citenamefont {Svensson},\ and\ \citenamefont {Suchalkin}}]{2010DonetskyAPL}%
  \BibitemOpen
  \bibfield  {author} {\bibinfo {author} {\bibfnamefont {D.}~\bibnamefont {Donetsky}}, \bibinfo {author} {\bibfnamefont {G.}~\bibnamefont {Belenky}}, \bibinfo {author} {\bibfnamefont {S.}~\bibnamefont {Svensson}},\ and\ \bibinfo {author} {\bibfnamefont {S.}~\bibnamefont {Suchalkin}},\ }\bibfield  {title} {\bibinfo {title} {Minority carrier lifetime in type-2 inas–gasb strained-layer superlattices and bulk hgcdte materials},\ }\href {https://doi.org/10.1063/1.3476352} {\bibfield  {journal} {\bibinfo  {journal} {Appl. Phys. Lett.}\ }\textbf {\bibinfo {volume} {97}},\ \bibinfo {pages} {052108} (\bibinfo {year} {2010})},\ \Eprint {https://arxiv.org/abs/https://doi.org/10.1063/1.3476352} {https://doi.org/10.1063/1.3476352} \BibitemShut {NoStop}%
\bibitem [{\citenamefont {DeWames}\ \emph {et~al.}(2019)\citenamefont {DeWames}, \citenamefont {Schuster}, \citenamefont {DeCuir},\ and\ \citenamefont {Dhar}}]{2019DewamesSPIE}%
  \BibitemOpen
  \bibfield  {author} {\bibinfo {author} {\bibfnamefont {R.~E.}\ \bibnamefont {DeWames}}, \bibinfo {author} {\bibfnamefont {J.}~\bibnamefont {Schuster}}, \bibinfo {author} {\bibfnamefont {E.~A.}\ \bibnamefont {DeCuir}},\ and\ \bibinfo {author} {\bibfnamefont {N.~K.}\ \bibnamefont {Dhar}},\ }\bibfield  {title} {\bibinfo {title} {{Recombination processes in InAs/InAsSb type II strained layer superlattice MWIR nBn detectors}},\ }in\ \href {https://doi.org/10.1117/12.2521907} {\emph {\bibinfo {booktitle} {Infrared Technology and Applications XLV}}},\ Vol.\ \bibinfo {volume} {11002},\ \bibinfo {editor} {edited by\ \bibinfo {editor} {\bibfnamefont {B.~F.}\ \bibnamefont {Andresen}}, \bibinfo {editor} {\bibfnamefont {G.~F.}\ \bibnamefont {Fulop}},\ and\ \bibinfo {editor} {\bibfnamefont {C.~M.}\ \bibnamefont {Hanson}}},\ \bibinfo {organization} {International Society for Optics and Photonics}\ (\bibinfo  {publisher} {SPIE},\ \bibinfo {year} {2019})\ p.\ \bibinfo {pages} {110020W}\BibitemShut {NoStop}%
\end{thebibliography}%

\end{document}